\documentclass{emulateapj}
\usepackage{natbib}
\usepackage{graphicx}
\usepackage{rotating}

\slugcomment{}
\shorttitle{Loop Top Emission}
\shortauthors{Sharma et al.}

\begin{document}

\title{On the Bright Loop Top Emission in Post Eruption Arcades}

\author{Rohit Sharma\altaffilmark{1}, Durgesh Tripathi\altaffilmark{2}, 
Hiroaki Isobe\altaffilmark{3} and Avyarthana Ghosh\altaffilmark{2,4}}

\altaffiltext{1}{National Centre for Radio Astrophysics, Post Bag 4, Ganeshkhind, Pune 411007, India}
\altaffiltext{2}{Inter-University Centre for Astronomy and Astrophysics, Post Bag 4, Ganeshkhind, Pune 411007, India}
\altaffiltext{3}{Graduate School of Advanced Integrated Studies in Human Survivability, Kyoto University, 
1 Yoshida-Nakaadachi-cho, Sakyo-ku, Kyoto 603-8306, Japan}

\altaffiltext{4}{IISER-Kolkata, Mohanpur - 741246, West Bengal, India}

\begin{abstract}
The observations of post eruption arcades (PEAs) in X-rays and EUV reveal strong localised brightenings at the loop top regions. The origin of these brightenings and their dynamics is not well understood to date. Here, we study the dynamics of PEAs using one-dimensional hydrodynamic modelling with the focus on the the understanding of the formation of localised brightening.Our findings suggest that these brightenings are the result of collisions between the counter-streaming chromospheric evaporation from both the foot points. We further perform forward modelling of the emission observed in simulated results in various spectral lines observed by the Extreme-Ultraviolet Imaging Telescope aboard Hinode. The forward modelled intensities in various spectral lines are in close agreement with a flare observed in December 17, 2006 by EIS. 
\end{abstract}

\section{Introduction}

Solar flares are possibly the brightest and most energetic bursts occurring in the atmosphere of the Sun. They release the high content of radiation and energetic particles into the interplanetary medium and play a key role in determining space-weather. Our understanding of the physics of solar flares is in a level that  there exists the "standard model", in which plasma ejection and magnetic reconnection play key roles \citep{shibata2011}, though there still remain many physical processes that are poorly understood. One such challenge is to explain the loop top emission observed in EUV and soft X-ray observations  of post flare loops or post eruption arcades \citep{priest_forbes2002}, which are considered to be one of the best proxies for Coronal Mass Ejections (CMEs) source regions \citep{tripathi2004}. However, no one-to-one correlation exists \citep[see e.g.,][]{ma2010,howard2013,chen2015,sun2015}.

One of the first observations of loop top emission in soft X-ray images was reported by \cite{acton1992}. Using the data taken from Soft X-ray Telescope (SXT) \citep{sxt1991} 
on board Yohkoh for 10 flares,  \cite{acton1992} found presence of compact X-ray sources at the loop tops, which were 
underneath the cusp-shape structures \citep{Forbes_action_1996}. Studying the flare that occurred on 13th January 1992, using the observations recorded by Yohkoh SXT and hard X-ray telescope 
(HXT) \citep{hxt1991,sxt1991}, \cite{masuda} discovered Hard X-ray sources at the loop top as well as the two footpoints of post 
flare loops. The HXR loop top observed by \cite{masuda} was located at the top of the SXR cusp detected in the SXT images.
The presence of cusp, and co-spatial HXR emission was attributed to magnetic reconnection processes \citep{Forbes_action_1996, priest_forbes2002}. However, there is no consensus on the origin of the soft X-ray and EUV brightening at the loop top.

\begin{figure}
\centering
\includegraphics[width=0.52\textwidth]{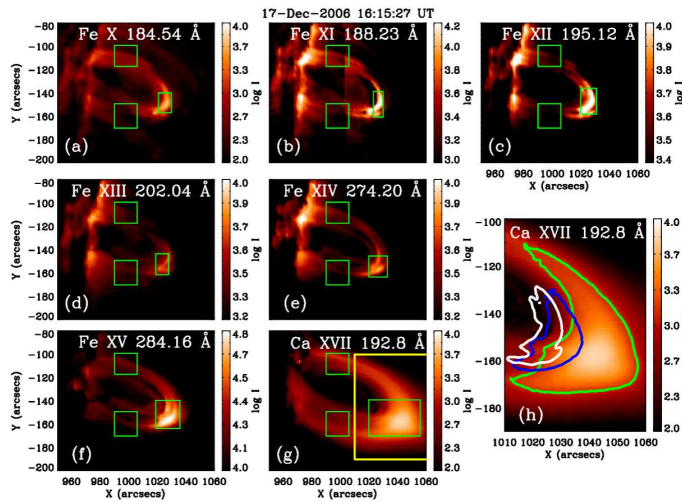}
\caption{ Intensity maps (a) to (g) for various spectral lines shown in Table \ref{table1}. The green boxes shows the area taken for the calculation of parameter brightening factor. Panel (h) is a zoom of yellow box of the \ion{Ca}{17} map panel (g). The Intensity contours in panel (h) are \ion{Ca}{17} (thick green), \ion{Fe}{15} (blue) and \ion{Fe}{12} (white).  Note that the contours are at 7$\%$ level of the maximum intensity in respective lines.\label{obsfigure}}
\end{figure}

With the launch of Hinode, it became possible to observe flares across a range of temperatures using the X-Ray Telescope (XRT)and the Extreme-ultraviolet Imaging Spectrometer (EIS).  \cite{Hara2006} reported the XRT and EIS observations of a limb flare on December~17,~2006 that show a clear cusp-structure and the loop-top brightening. With the EIS instrument, \cite{Hara2006} could investigate the flare dynamics across a range of temperature. Some iron and calcium spectral lines having formation temperatures from $\log\, T = 6.0$ to $6.7$ is shown in Table \ref{table1}. In Fig. \ref{obsfigure}, the panel (a) to (g) shows  intensity maps for 6 iron lines \ion{Fe}{10} to \ion{Fe}{15} and \ion{Ca}{17}. In the same figure, the panel (h) shows zoomed maps of loop top region and contours of \ion{Fe}{15} and \ion{Fe}{12}. As can be depicted from the Fig. \ref{obsfigure}, panel (h) and Table \ref{table1}, hotter plasma is located at higher coronal heights (i.e., in the outer loops) and cooler plasma is at lower coronal heights (in the inner loops). This is the well-known characteristics since the Yohkoh observations \citep{tsuneta1996} that is consistent with the "standard model" with magnetic reconnection \citep{yokoyama2001}. A number of authors have shown that the loop top brightening is not only seen in SXR, but also in EUV observations using narrow band imager \cite[see e.g.,][]{golub1999,warren1999,warren2000,warren2001,doschek2006} obtained by the Transition Region and Coronal Explorer \citep[TRACE;][]{Handy1999}. However, \cite{Hara2006}, for the first time made a clear distinction using the spectroscopic observations by Hinode/EIS.

There have been numerous attempts to explain the above described phenomenon using various approaches such as hydrodynamic modelling of multiple loops \citep{hori1997}, 1-D hydrodynamic modelling in soft X-ray and EUV for SXT and TRACE respectively \citep{reeves2007} and retracting loops forming gas-dynamic shock \citep{longcope2009,longcope2011}. However, there is no consensus on the reason behind the observed loop top brightenings in SXR and EUV.

The loop top brightening can be quantified by taking the ratios of the intensities at the loop top region and arm regions of the coronal loop shown in the green boxes in Fig. \ref{obsfigure} and is called brightening factor ($\beta_{obs}$). The intensity in the arm regions is calculated by averaging over two coronal arm regions (Fig. \ref{obsfigure}). Table \ref{table1} shows the intensity values and $\beta_{obs}$ at the loop top for various spectral lines. $\beta_{obs}$ varies in the range of 1.5 to 3.7. Note that the \ion{Fe}{15} is the brightest spectral line.

In order to explain the bright knots observed in EUV post flares loops with various instruments \citep{cheng1980,dere1997,widing1984,golub1999,warren2000}, 
\cite{patsourakos2004} performed a 1D hydrodynamic modelling, similar to the calculations performed by Antiochos (1980).  It was found that the formation of these bright knots can be explained by spatially localised heating that may be occurring during the decay phase of the flares rather than spatially uniform heating. 

In this study, we take a similar approach to study the evolution of the loop top emission in post flare loops observed in soft X-ray 
and EUV observations. We further perform forward modelling and compare our simulation results with observations provided by EIS for the
flare that was studied by \cite{Hara2006}. The rest of the paper is structures as follows. The \S~\ref{model} we provide the details of the 
numerical model along with initial and boundary conditions. The section \S~\ref{results} summarises the simulation results. The role of 
energetics and evolution of density, temperature and velocity is discussed for various heating inputs. Next \S~\ref{foward_modelling}, deals with the 
forward modelling of the loop tops in EIS spectral wavebands. Lastly \S~\ref{summary}, discusses the results and conclusions.

%%%--------------------------------------------------------
%%%-----------------------------------------------------
%%------------------------------------------------------
\begin{table}[htbp]
	\centering
\caption{The observation parameters for various spectral lines for 17th December 2006 flare.}
\begin{tabular}{ccccc}
\tableline\tableline
\vbox{\hbox{\strut Spectral }\hbox{\strut lines}} & \vbox{\hbox{\strut Wavelength }\hbox{\strut (in \AA)}} & \vbox{\hbox{\strut Peak} \hbox{\strut (log T)}} & \vbox{\hbox{\strut Loop top Intensity }\hbox{(ergs/cm$^{2}$/sec/str)}} & \vbox{\hbox{\strut Brightening }\hbox{factor ($\beta_{obs}$)}}   \\
\tableline
Fe X & 184.54&6.00& 5.1 $\times$ 10$^{3}$ &3.67\\
Fe XI & 188.23&6.08& 2.6 $\times$ 10$^{4}$ &2.32\\
Fe XII & 195.12&6.10& 1.2 $\times$ 10$^{4}$ &2.91\\
Fe XIII & 202.04&6.15& 5.7 $\times$ 10$^{3}$ &1.54\\
Fe XIV & 274.2&6.25&7.6 $\times$ 10$^{3}$&1.65\\
Fe XV & 284.16&6.30 &6.5 $\times$ 10$^{4}$&2.18\\
Ca XVII &192.82&6.70& 6.8 $\times$ 10$^{3}$&2.70\\
\tableline
\end{tabular}
\label{table1}
\end{table}

%%----------------------------------------------------------------------
\section{Numerical Model} \label{model}

\begin{figure}[htbp]
\centering
\includegraphics[width=0.4\textwidth]{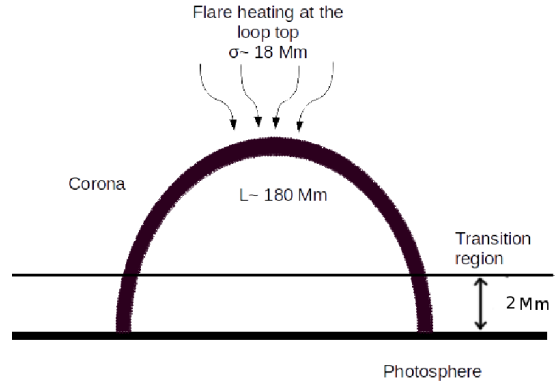}
\caption{Schematic of the 1-D coronal loop, indicating the location of localised flare heating. The total length of the loop (L) is 180 Mm. The transition region location is kept at 2 Mm. \label{schematic}}
\end{figure}

 We simplify the problem to one dimensional hydrodynamics of a semi-circular loop with constant cross section
 as shown in Figure~\ref{schematic}. To make the simulation
results observationally more viable, we have taken the parameters such as length, approximate energy deposition based on 
the parameters measured for the C-class flare observed on 17th~December~2006, which was studied by \cite{Hara2006}. The length of the loop is $\sim$~180~Mm.

%%----------------------------------------------------------------------
%%----------------------------------------------------------------------

In this simulations we solve the standard hydrodynamic equations, which are described in Equations~\ref{cont}, 
\ref{moment} \& \ref{eng} with an equation of state of ideal gas in c.g.s units.

%%----------------------------------------------------------------------
\begin{equation}\label{cont}
\frac{\partial \rho}{\partial t} + \frac{\partial}{\partial s}(\rho v)= 0
\end{equation}
%%----------------------------------------------------------------------

%%----------------------------------------------------------------------
\begin{equation}\label{moment}
\frac{\partial}{\partial t}(\rho v) + \frac{\partial}{\partial s}(\rho v^{2}) = -\rho g_{||}-\frac{\partial P}{\partial s}
\end{equation}
%%----------------------------------------------------------------------

%%----------------------------------------------------------------------
\begin{equation}\label{eng}
\frac{\partial E}{\partial t} + \frac{\partial}{\partial s} [(E+P)v]+ \rho g_{||} v = \frac{\partial}{\partial s}(\kappa_{||} \frac{\partial T}{\partial s}) -R +H
\end{equation}
%%----------------------------------------------------------------------

\noindent where,
%%----------------------------------------------------------------------
\begin{equation}\label{ideal}
P=n k_{B}T
\end{equation}
%%----------------------------------------------------------------------
%%----------------------------------------------------------------------
\begin{equation}\label{gamma}
E= \frac{1}{2} \rho v^{2} + \frac{P}{\gamma -1}
\end{equation} 
%%----------------------------------------------------------------------

Here $\rho$ is the total mass density, $v$ is the fluid velocity, $P$ is the total gas pressure, $E$ is the sum of
the kinetic energy and the internal energy per unit volume, $T$ is the plasma temperature and $s$ is the length along the loop from the left footpoint. We have assumed a monoatomic ideal gas with specific heat capacity $\gamma=5/3$. The gravity 
$g_{||}$ is $g_{o}\ cos(\frac{h}{r})$, where \textsl{h} is the height from the photosphere, \textsl{$r$} is the 
radius of curvature of the loop and 
 $g_{o}$ is 2.7~$\times$~10$^{4}$~cm/sec$^{2}$.

The heat conduction is governed by temperature gradients across the loop with Spitzer conductivity given by

%%----------------------------------------------------------------------
\begin{equation}\label{cond}
\kappa_{||}=\kappa_{0} T^{5/2}
\end{equation} 
%%----------------------------------------------------------------------

\noindent where $\kappa_{0}=9\times$~10$^{-7}$. Note that all the units here are in CGS.

In equation~\ref{eng}, H is a heating function \textsl{H(s, t)} that consists of a static heating $H_{s}$ 
and an flare heating term $H_{f}$. The functional form of H is given as:

%%----------------------------------------------------------------------
\begin{equation}
	H(s,t)=H_{s}(s)+H_{f}(s,t)
\end{equation}
%%----------------------------------------------------------------------

The static heating is kept constant to maintain the corona at 2~MK. The flare heating term provides the heating 
due to flare to which the plasma responds and produces the dynamics in the loop. The functional form of the flare 
heating is given by Equation~\ref{heat}. In this study we have considered the flare heating to have a Gaussian shape
with a width of $\sigma \sim$~18~Mm. The mean of the Gaussian is placed at the loop top $s_{top} \sim$~90~Mm. 

%%----------------------------------------------------------------------
\begin{equation}\label{heat}
H_{f}(s,t)=q(t) \frac{H_{o}}{\sqrt{2 \pi} \sigma} exp \Big( -\frac{(s-s_{top})^{2}}{2 \sigma^2}\Big) \ (ergs \ cm^{-2})
\end{equation}
%%----------------------------------------------------------------------

\noindent where q(t) is the binary switch for the flare heating, i.e., if q(t) is 0 then heating is off and if 1 then 
heating is on. $H_{o}$ defines the heating strength, which is kept as a variable. By varying $H_{o}$, we can study the
dynamics of plasma in flare loops with varying heating strengths. 

\textsl{R} in Eqn.~\ref{eng} is the radiative cooling, which is defined as

%%----------------------------------------------------------------------
\begin{equation}
R(s,t)=\frac{1}{2}~n_{e}^{2}~Q(T)
\end{equation}
%%----------------------------------------------------------------------

\noindent where $Q(T)$ is the radiative loss function for optically thin plasma and is approximated through various
power law functions  in different temperature bands in the format $Q(T)~=~\chi~T^{\alpha}$ ($ergs~s^{-1}~cm^{-3}$) and
taken directly from \cite{hori1997}.

%%----------------------------------------------------------------------
\begin{figure}
\centering
\includegraphics[width=0.4\textwidth]{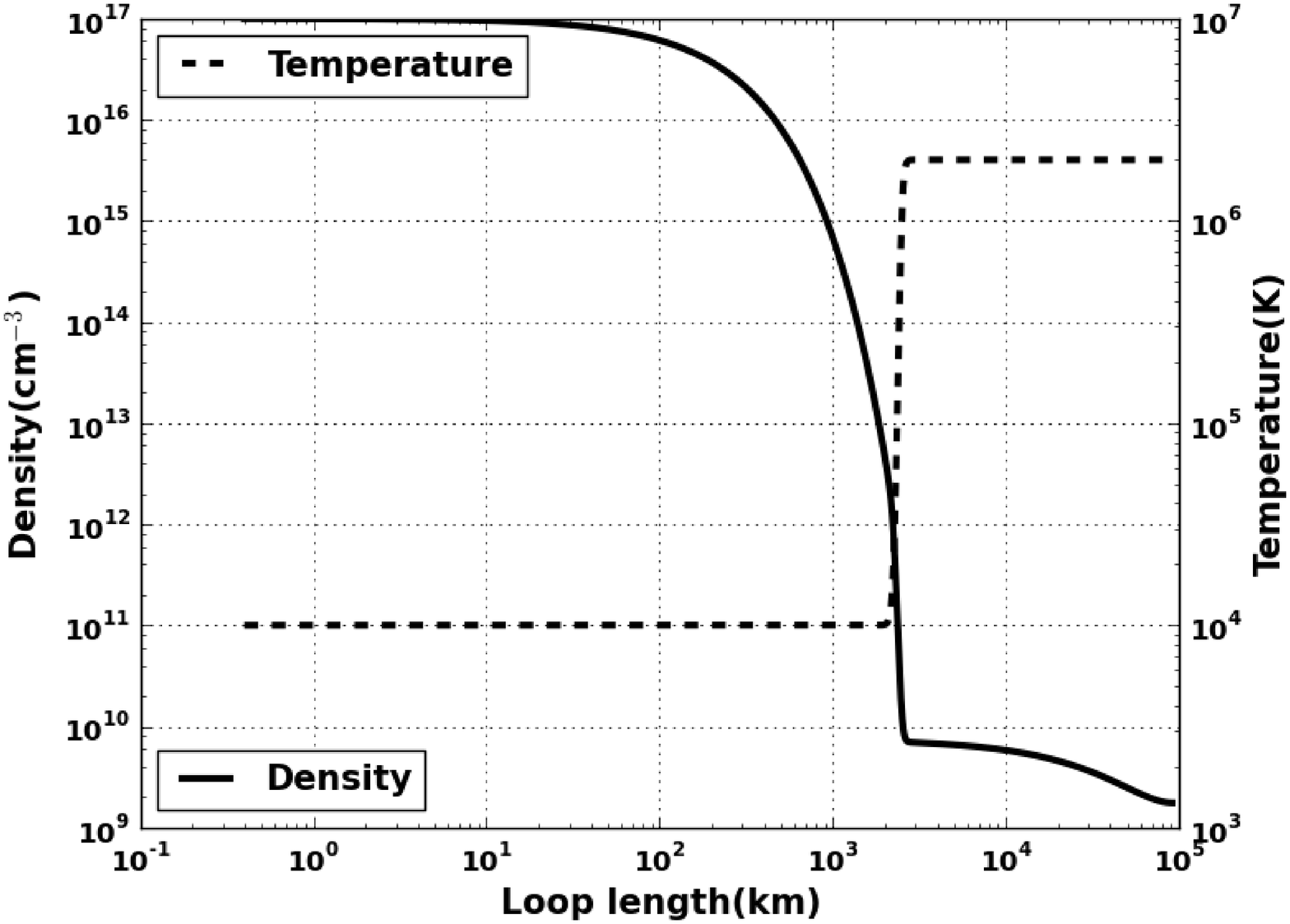}
\caption{Initial temperature and density profiles for the half loop.\label{initial}}
\end{figure}
%%----------------------------------------------------------------------

Figure~\ref{initial} shows the initial conditions for our simulation. We have started the simulations with realistic 
initial conditions in terms of temperature and density profile for the solar atmosphere. The system is set up in such as way 
that the loop is in hydrostatic equilibrium, i.e., static heating fully compensate the radiative cooling energy. The initial 
temperature is kept at 2~MK because we are mostly interested in active region post flare loops which are best seen in \ion{Fe}{12} 
lines forming at 2~MK. The density is then computed along the loop using the equations of hydrostatic equilibrium. 

We have also considered the loop to be symmetric across its length.
Considering this geometry leads us to solve the equations only for half of the loop (from one footpoint to the loop top). The solution
of the other part will then become the mirror image. The boundary conditions are set up so that there are no gradient
in temperature and density at the loop top. Mathematically the boundary condition can be written as 
$(\frac{\partial n}{\partial s})_{top}=0$, $(\frac{\partial T}{\partial s})_{top}=0$ and $v_{top}=0$.

%%----------------------------------------------------------------------
\section{Results }\label{results}
%%----------------------------------------------------------------------

\subsection{Overview}
We have performed simulations for six different external heating (H$_f$) applied at the loop top. The heating input to the loop will be 2~$\times$~10$^{9}$, 2~$\times$~10$^{10}$, 4~$\times$~10$^{10}$, 8~$\times$~10$^{10}$, 1.2~$\times$~10$^{11}$, 2.0~$\times$~10$^{11}$ ergs~cm$^{-2}$ uniformly in a time of 100~sec each. These energy values corresponds to the energy of a C-class flare ($\sim$ 10$^{29}$ ergs) distributed over 18~Mm length. We note that the flare observed by \cite{Hara2006} is a C-class flare. A few different values of heating strengths are chosen so as to cover two orders of magnitude centring at C-class flares strengths.

One of the first parameters which needs to be verified is the principle of energy conservation throughout the simulations. We note that before the deposition of any external heat, the system is in hydrostatic equilibrium i.e. the static heating in the system balances out the radiative cooling. After the heat deposition, we have carefully looked at the different form of energy available in system and their evolution.

%%----------------------------------------------------------------------
\begin{figure}
\centering
\includegraphics[width=0.4\textwidth]{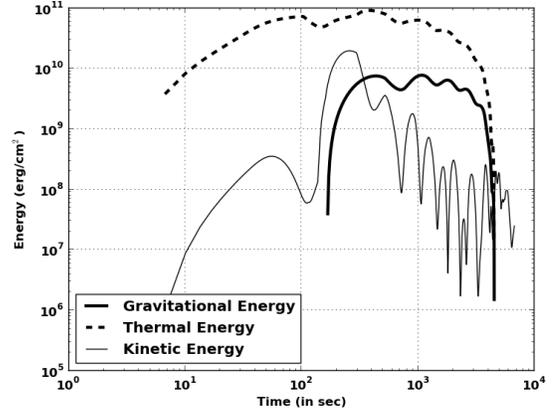}
\caption{Comparison of the different energies in the corona with time. The gravitational($\Delta E_{GE}$) and thermal energy($\Delta E_{TE}$) shown is the variation from intial gravitational and thermal energy of the loop respectively. } \label{energy}
\end{figure}
%%----------------------------------------------------------------------

First we discuss the energetics and dynamics of the loop evolution for the highest heating case i.e. 
2.0~$\times$~10$^{11}$~ergs~cm$^{-2}$ in detail. We note that same line of arguments will follow for the other heating
scenarios. We calculate total thermal ($E_{TE}$), kinetic ($E_{KE}$) and gravitational energy ($E_{GE}$) in the corona. The coronal part of the loop is defined as the portion of the loop above than 2 Mm. Figure~\ref{energy} shows the temporal evolution of the changes in 
the total thermal energy ($\Delta E_{TE}$), total kinetic energy ($\Delta E_{KE}$) and total gravitational energy ($\Delta E_{GE}$) in the coronal part of the loop with respect to that at t=0. For energy component $E$, $\Delta E = E(t) - E(t=0)$. There is no total kinetic energy in the loop at t=0 as $v(s)=0$, so $\Delta E_{KE} = E_{KE}$.

As can be seen from 
the Figure~\ref{energy}, the thermal energy ($\Delta E_{TE}$) starts to increase and attains a maximum at time $\sim$100~sec.
This is consistent with fact that the external heat was supplied only for the first 100~sec. Thereafter, the thermal energy ($\Delta E_{TE}$) shows small fluctuations $\sim 10\%$  and declines sharply after $\sim$ 4000~sec. Similarly, kinetic energy($\Delta E_{KE}$) also increases and attains a maximum at around $\sim$60~sec. It reaches minima at time $\sim$ 102~sec and rises sharply reaching second maxima at time $\sim$ 300~sec. Thereafter kinetic energy fluctuates and the level of fluctuation in the kinetic energy is $\sim 50\%$. Also, $\Delta E_{GE}$ is nil for first 100~sec. It starts to increase sharply at $\sim$105~sec and attains a maximum at 300~sec and later fluctuate at $\sim 10\%$ level. Afterwards, it falls sharply at around $\sim$ 3500~sec. 

\begin{figure*}
\begin{center}
\begin{tabular}{cc}
\includegraphics[scale=0.20]{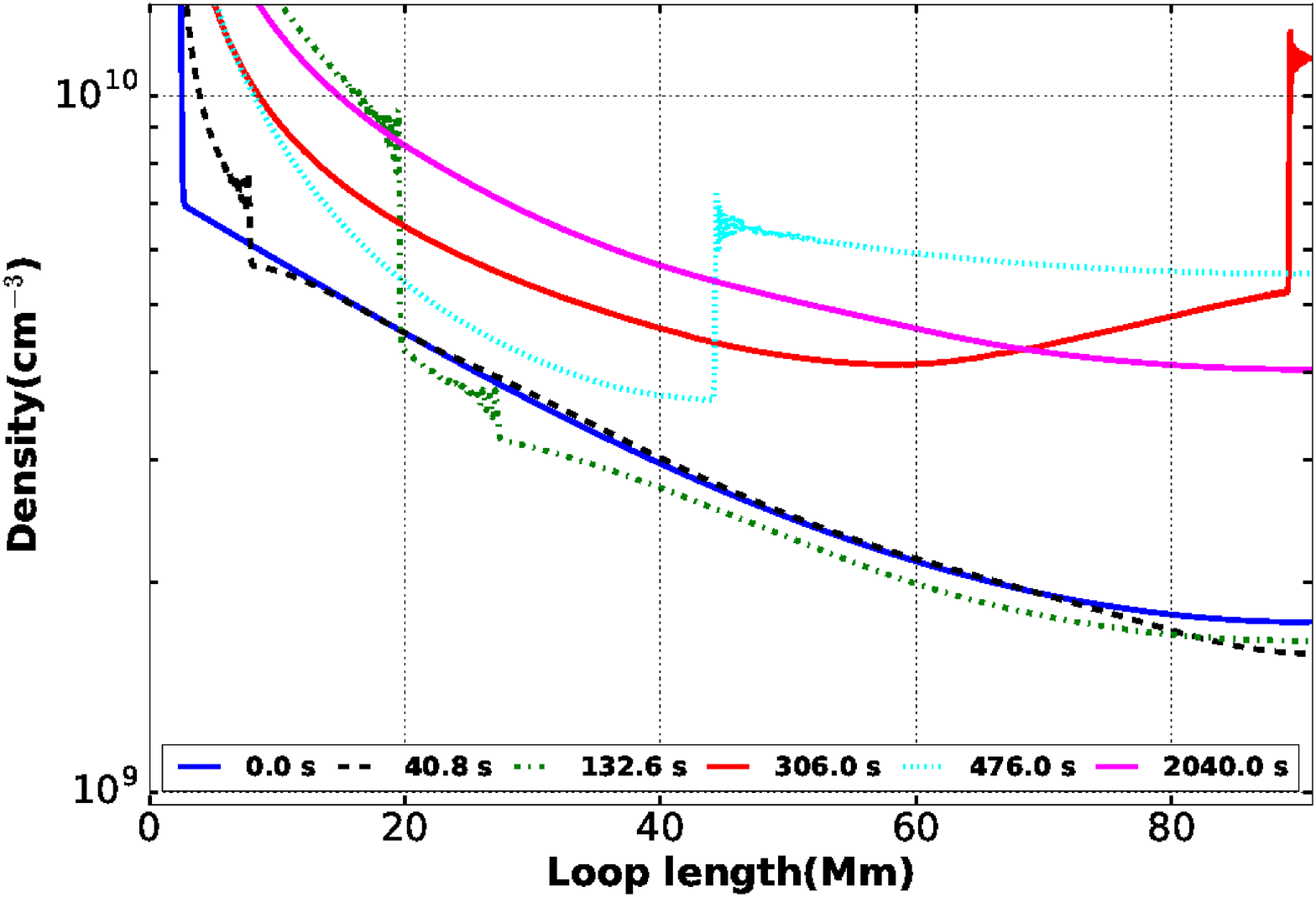}
  &
\includegraphics[scale=0.20]{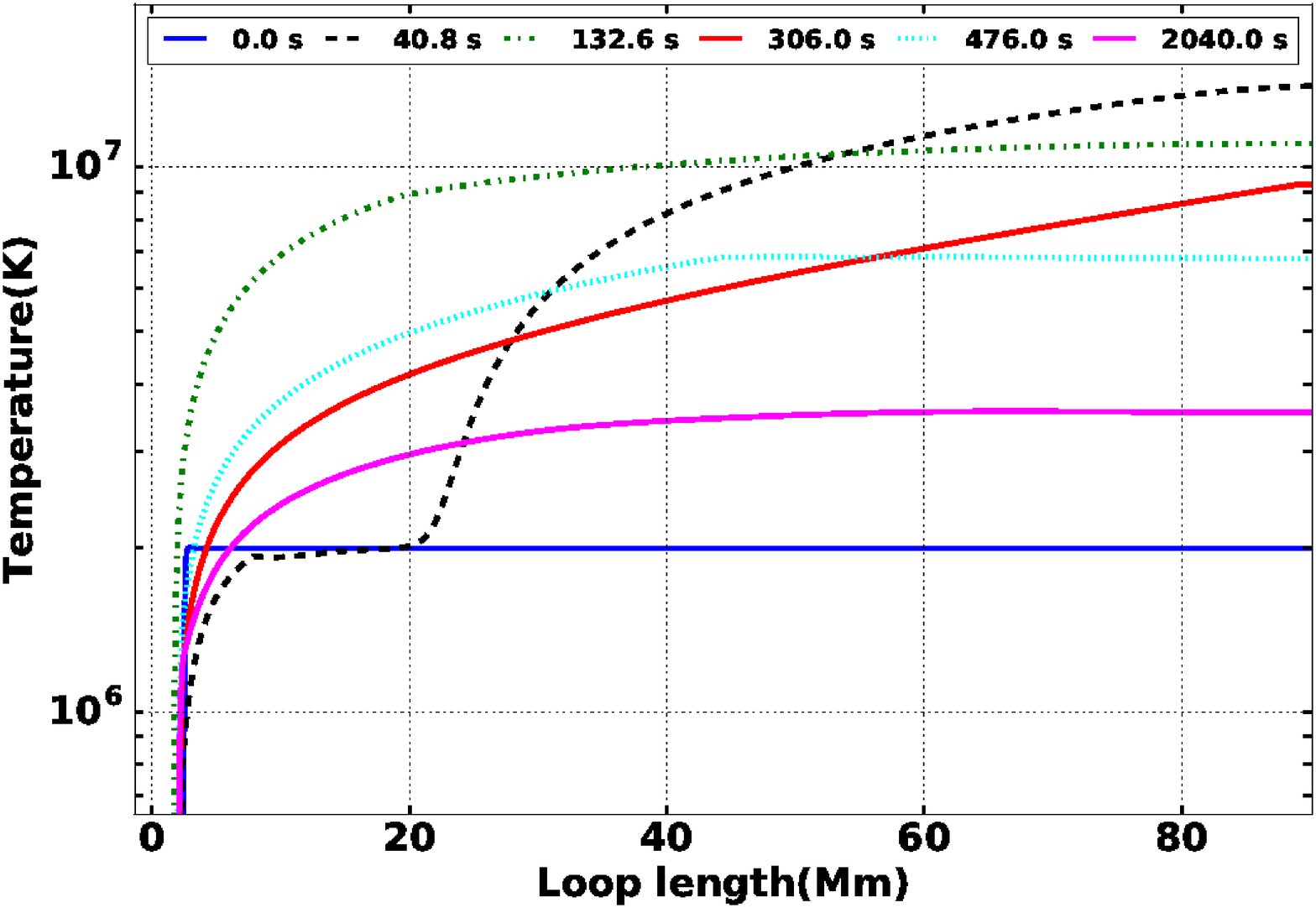}
 \\
  (a) Density&(b) Temperature\\
\includegraphics[scale=0.20]{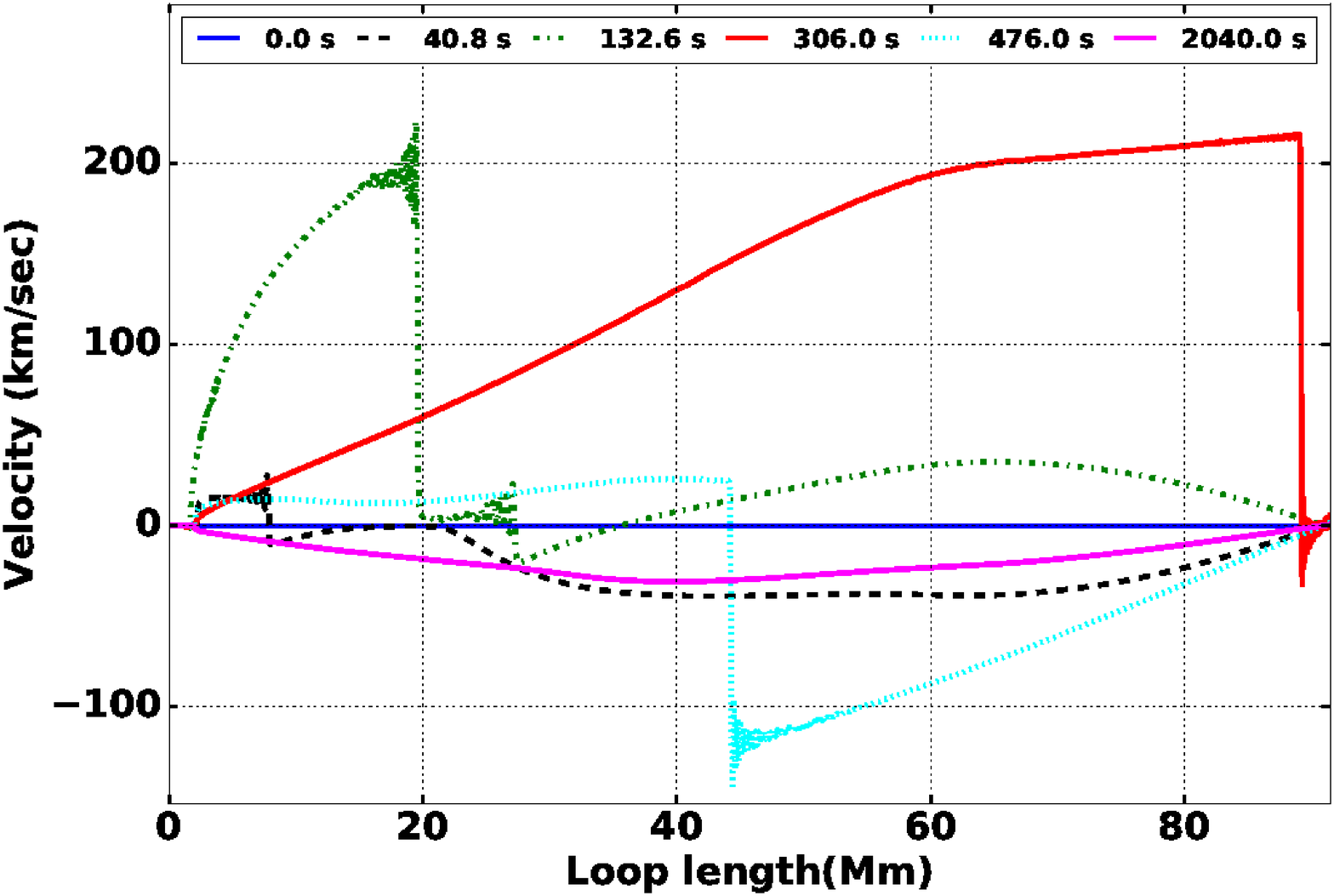}
&
\includegraphics[scale=0.20]{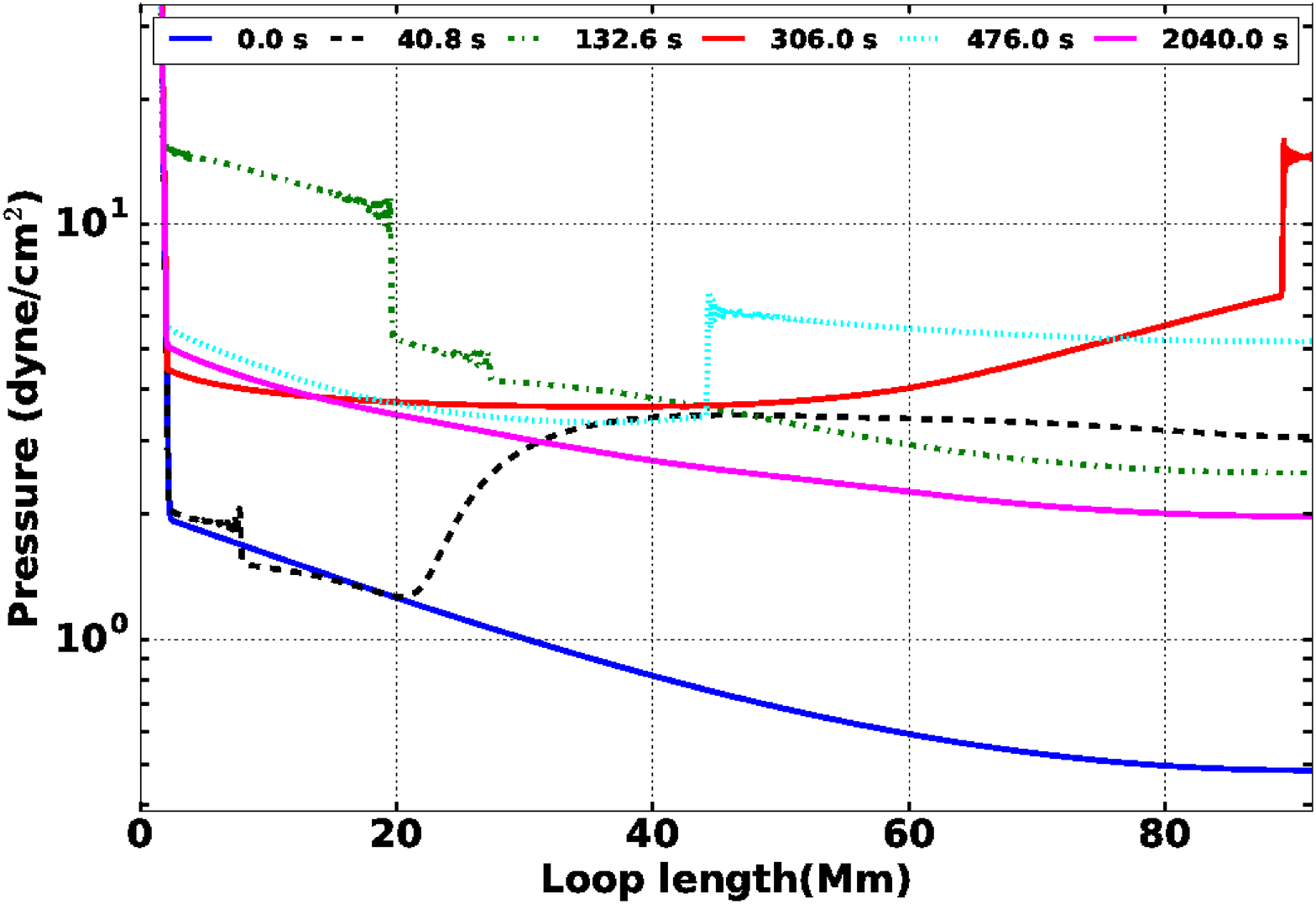}
 \\
(c) Velocity &(d) Pressure

\end{tabular}
\end{center}
\caption{The panel shows the plots of density, temperature, velocity and pressure at various times for input heating of 
2~$\times$~11$^{10}$~ergs~cm$^{-2}$. }
\label{specific_case}
\end{figure*}

Figure~\ref{specific_case} plots the time evolution of density (panel~a), temperature (panel~b), velocity (panel~c) and pressure(panel~d). Note that the plots display the above said parameters only for the half of the loop. The initial profile for all parameters is plotted at t=0. At t=40~sec, the temperature (Fig.~\ref{specific_case}, panel (b)) of the loop top rises to 10~MK and temperature profiles develops a spatially broader profile. The velocity profile (Fig.~\ref{specific_case}, panel (c)) is roughly uniformly negative in higher corona from 25~Mm to 80~Mm and shows small bump near length $\sim$~8~Mm. 

At t~=~132~sec, the coronal temperature profile flattens to $\sim$~10~MK. %By that time the conduction front reaches the chromosphere, increases the gas pressure of the chromosphere and drives the upward evaporation flow.
The velocity profile develops strong positive profile reaching speeds of 200~km~sec$^{-1}$. The density develops a steep gradient near 20 Mm and pressure profile now shows a negative dip near the same location. Since the flare heating stops at t=100~s, the coronal temperature dips (t=306~s, panel(b)). At 306~sec, the density and pressure profiles (Fig.~\ref{specific_case}, panel (a) \& (d)) shows a sharp increase at the loop top and velocity (panel(c)) shows large positive profile near loop top at length $\sim$88 Mm. The sharp enhancement in density propagate towards bottom of the loop (Fig.~\ref{specific_case}, panel (a) \& (d), t=476~s) with a velocity (panel(c), t=476~s) of 120~km/sec. At very large times t~=~2040~sec, the loop cools down to few million K. However, the density and pressure profiles quietens and settles. The velocity profile becomes spatially flatter and negative.   

The evolution of energy shown in Figure~\ref{energy} as well as the other physical parameters shown in Figure~\ref{specific_case} 
can be described as follows. As can be seen in Figure~\ref{specific_case}~panel b, the external heat input increases the 
temperature (broader bump at t=40~s), that in turn increases the thermal energy that is reflected in Figure~\ref{energy} (dashed line).
The increase in thermal energy($\Delta E_{TE}$) creates a pressure gradient from the loop top to the loop bottom (see panel d, t=40~s) that invokes 
a downward velocity flow or conduction front (from loop top to loop footpoint) with a speed of 42~km/sec 
uniformly throughout the corona as shown in panel c (t=40~s).
Here, we describe negative velocity as the flow from top to bottom. Due to the downflowing plasma, the kinetic energy($\Delta E_{KE}$) in system 
increases, which is also reflected in Figure~\ref{energy} (thin solid line).

The downward conduction front (downflow) produced due to the pressure gradient hits the chromosphere at $\sim$120-130~sec and 
produces fast upflows (see Fig. \ref{specific_case} panel c t=132~sec), a phenomena known as "chromospheric evaporation". Due to 
the chromospheric evaporation, the thermal energy($\Delta E_{TE}$), kinetic energy($\Delta E_{KE}$), as well as the gravitational ($\Delta E_{GE}$) show an increase. The velocity of the 
chromospheric evaporation reaches up to $\sim$220~km~sec$^{-1}$. Since 
the loop is considered to be symmetric and external heat input is at the center of the loop, the evaporation flow has the same speed from both the foot points. The evaporation flows from the two points collide at the loop top. We refer to it as primary collision of the evaporation flow and it leads to a sharp increment in density at loop top (see Fig. \ref{specific_case} (a) t=306~sec) as well as thermal energy 
(see Figure~\ref{energy} dashed line). After the primary collision, the mean kinetic energy($\Delta E_{KE}$) starts to decline and show oscillatory behaviour. This 
is suggestive of the fact that there may be secondary, tertiary multiple downflows and upflows. The gravitational energy($\Delta E_{GE}$), however, remains rather 
constant until t=3500~sec, suggesting no significant movement of matter. Therefore, it is plausible to conclude that the secondary and other downflows/upflows replicated in kinetic energy curve in Fig. \ref{energy} are weak.

At later times ($\sim$ 2000 sec and afterwards), the density in the corona of the loop is uniformly higher compared to initial 
density profile by a factor of 3.  The gravitational energy ($\Delta E_{GE}$) remains flat till 2000~sec with very small variation (see 
Fig.~\ref{energy}). Both the thermal energy ($\Delta E_{TE}$) as well as gravitational energy($\Delta E_{GE}$) show steep decline afterwards. This is suggestive of
strong cooling and draining of plasma from the corona to the chromosphere. 

%----------------------------------------------------------------------
%----------------------------------------------------------------------

%%----------------------------------------------------------------------
\subsection{Dynamics of the loop for various input heats}\label{dynamics}
%%----------------------------------------------------------------------

\begin{figure*}
\begin{center}
\begin{tabular}{cc}
\includegraphics[scale=0.22]{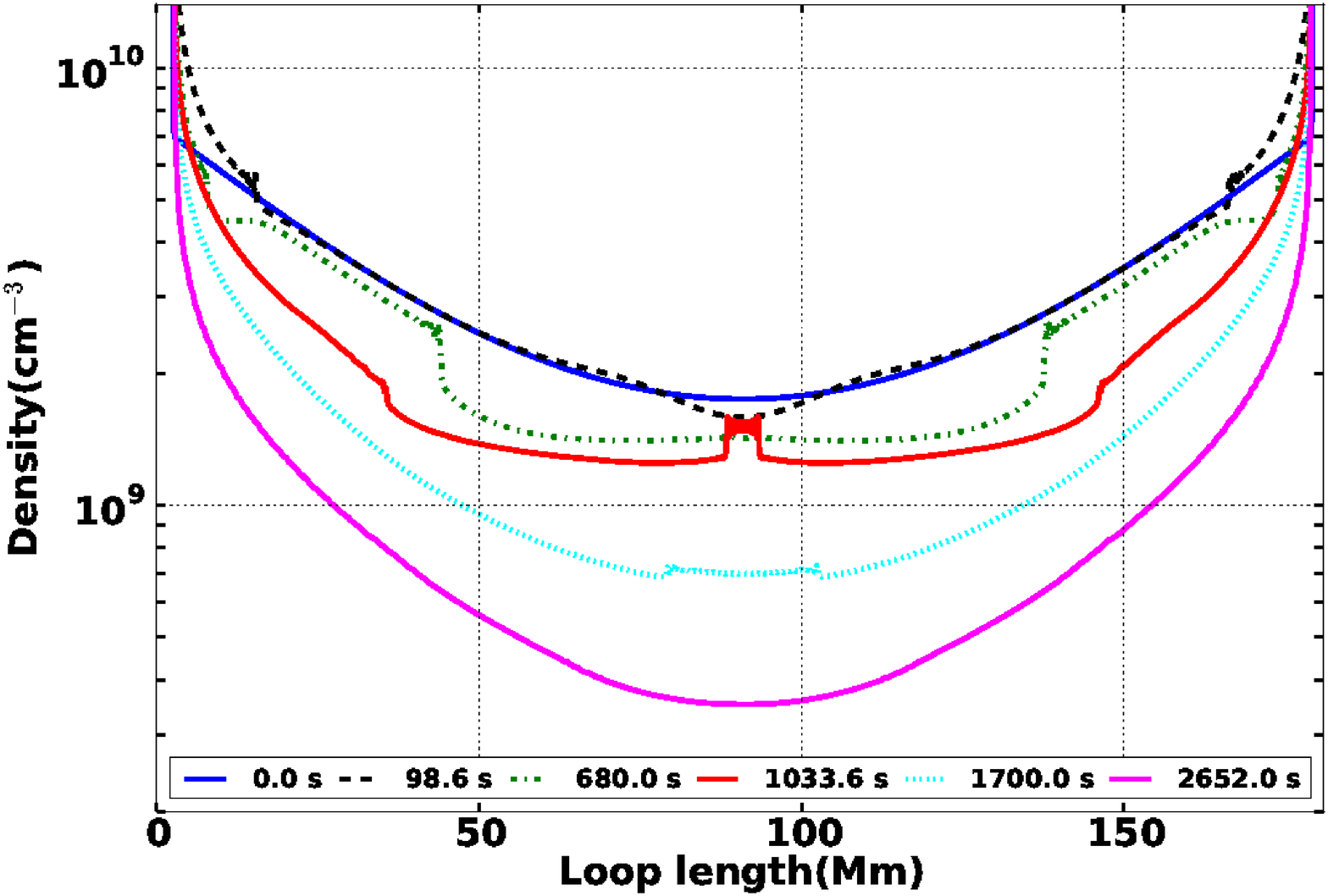}
  &
\includegraphics[scale=0.22]{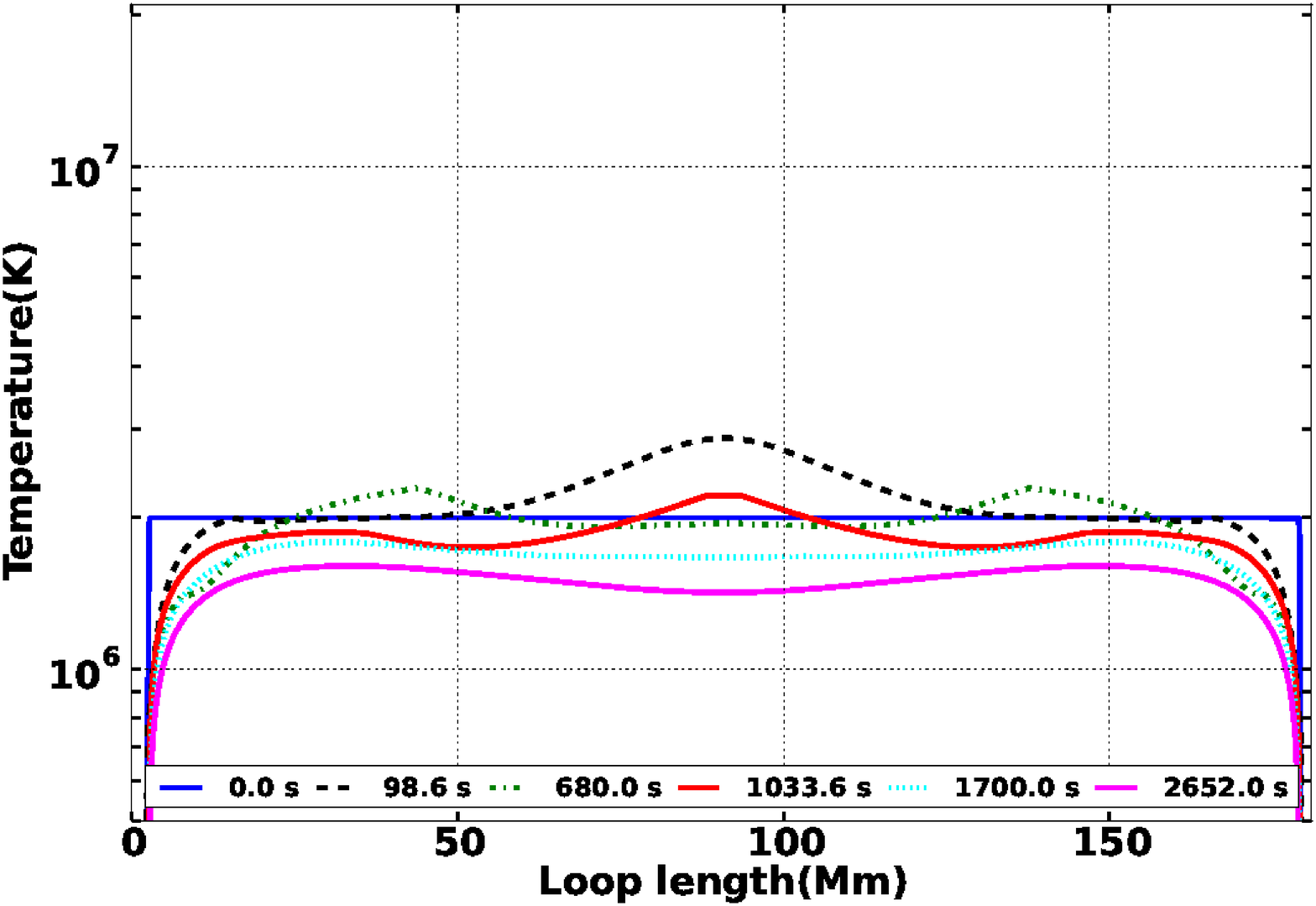}
 
 \\ (a) Density for 2 $\times$ 10$^{9}$ \ ergs/cm$^{2}$&(b) Temperature for 2 $\times$ 10$^{9}$ \ ergs/cm$^{2}$\\
\includegraphics[scale=0.22]{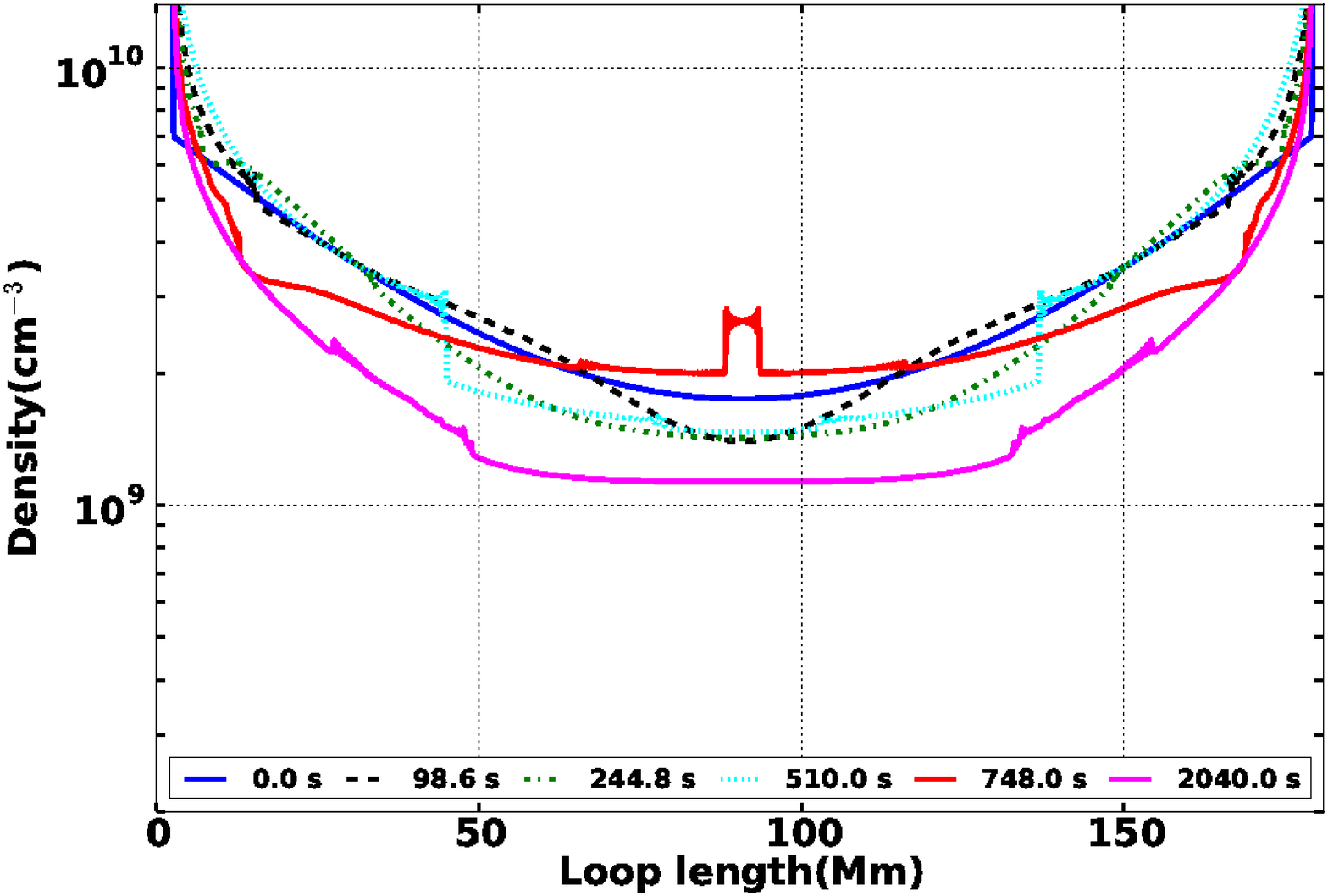}
 &
 \includegraphics[scale=0.22]{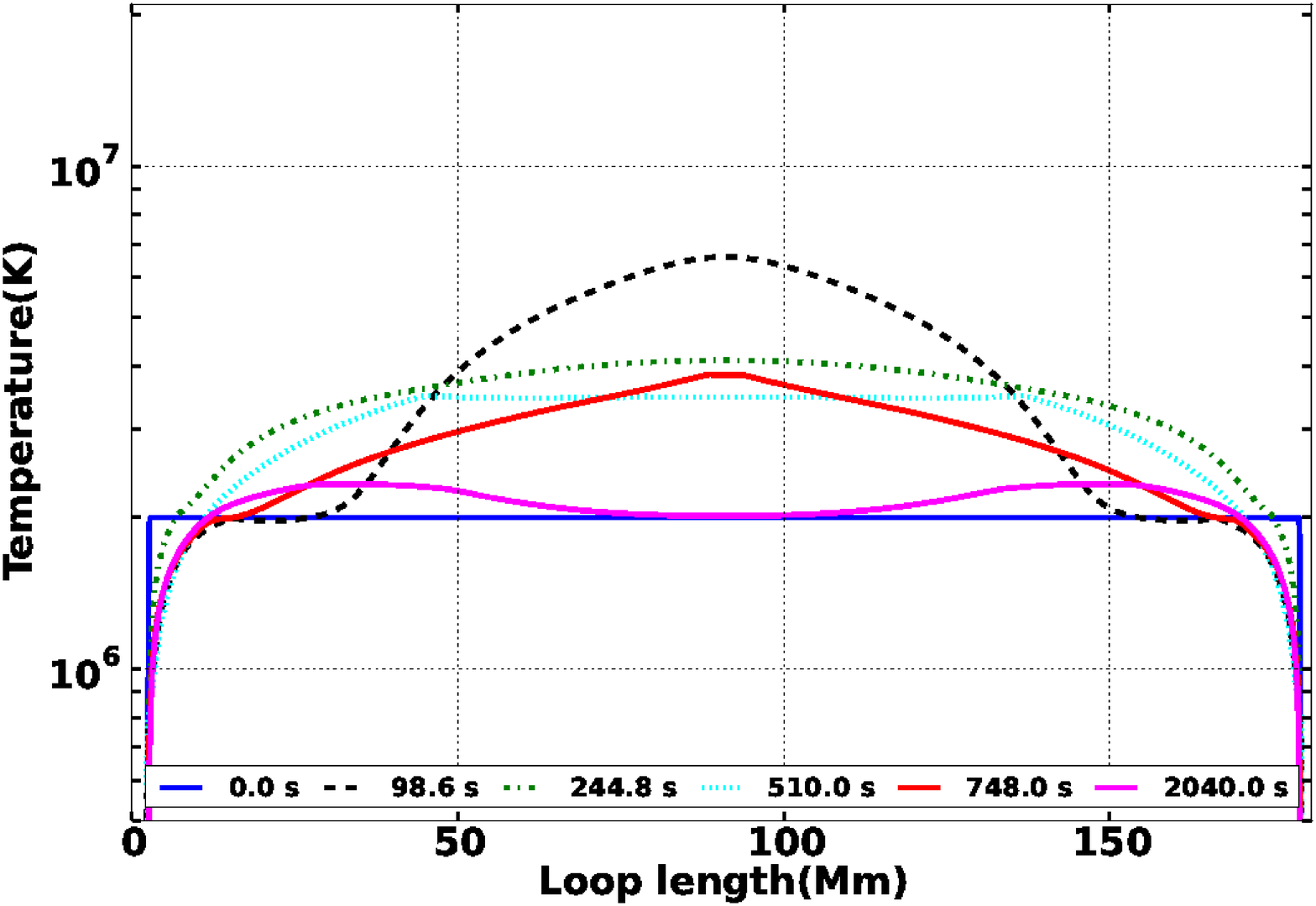}
 \\ (c) Density for 2 $\times$ 10$^{10}$ \ ergs/cm$^{2}$&(d)Temperature for 2 $\times$ 10$^{10}$ \ ergs/cm$^{2}$\\
 \includegraphics[scale=0.22]{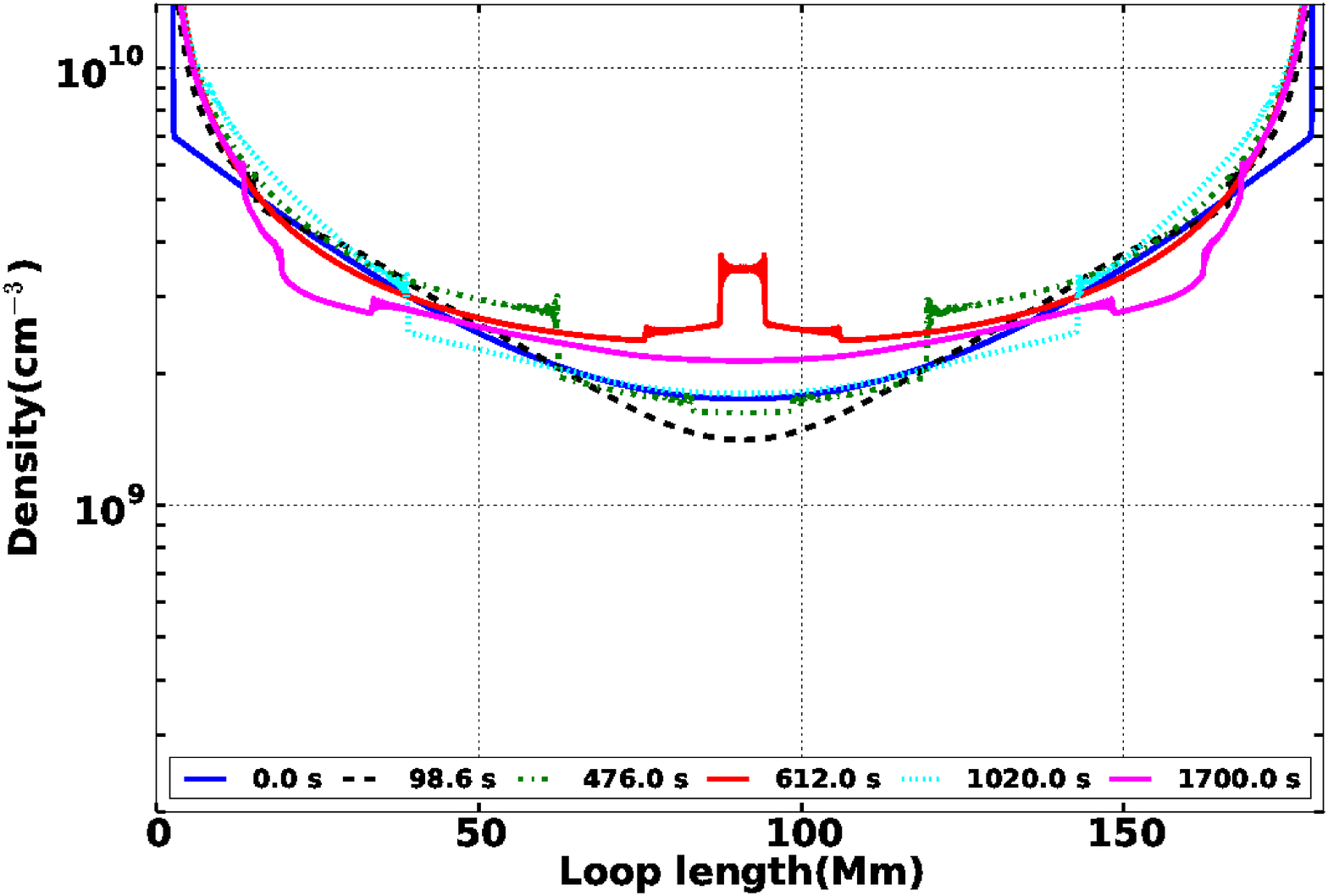}
 &
 \includegraphics[scale=0.22]{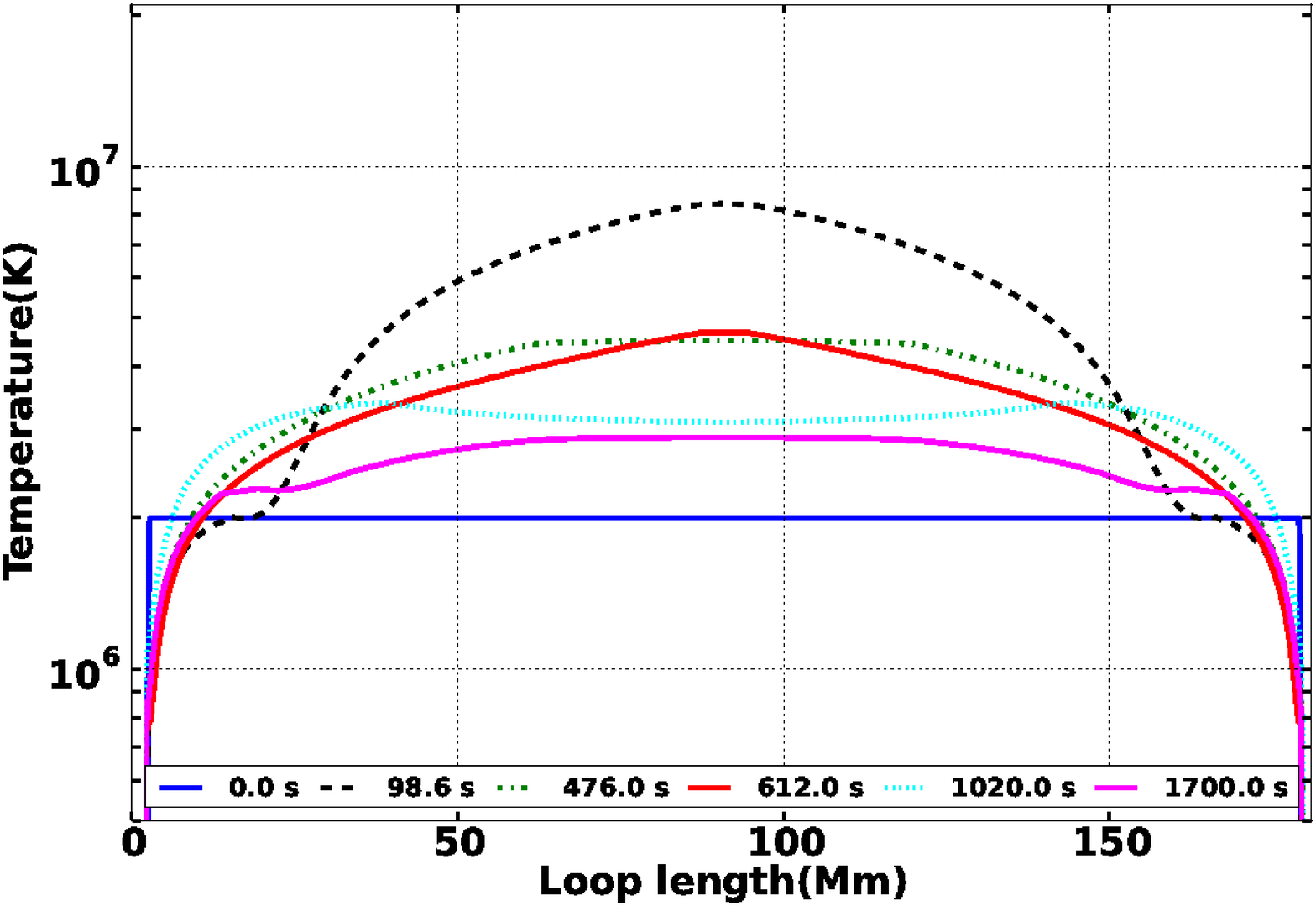}
 \\ (e) Density for 4 $\times$ 10$^{10}$ \ ergs/cm$^{2}$&(f)Temperature for 4 $\times$ 10$^{10}$ \ ergs/cm$^{2}$
\end{tabular}
\end{center}
\caption{Density and Temperature evolution for various heating strengths.\label{density_temp1}}
\end{figure*}

\begin{figure*}
\begin{center}
\begin{tabular}{cc}

\includegraphics[scale=0.22]{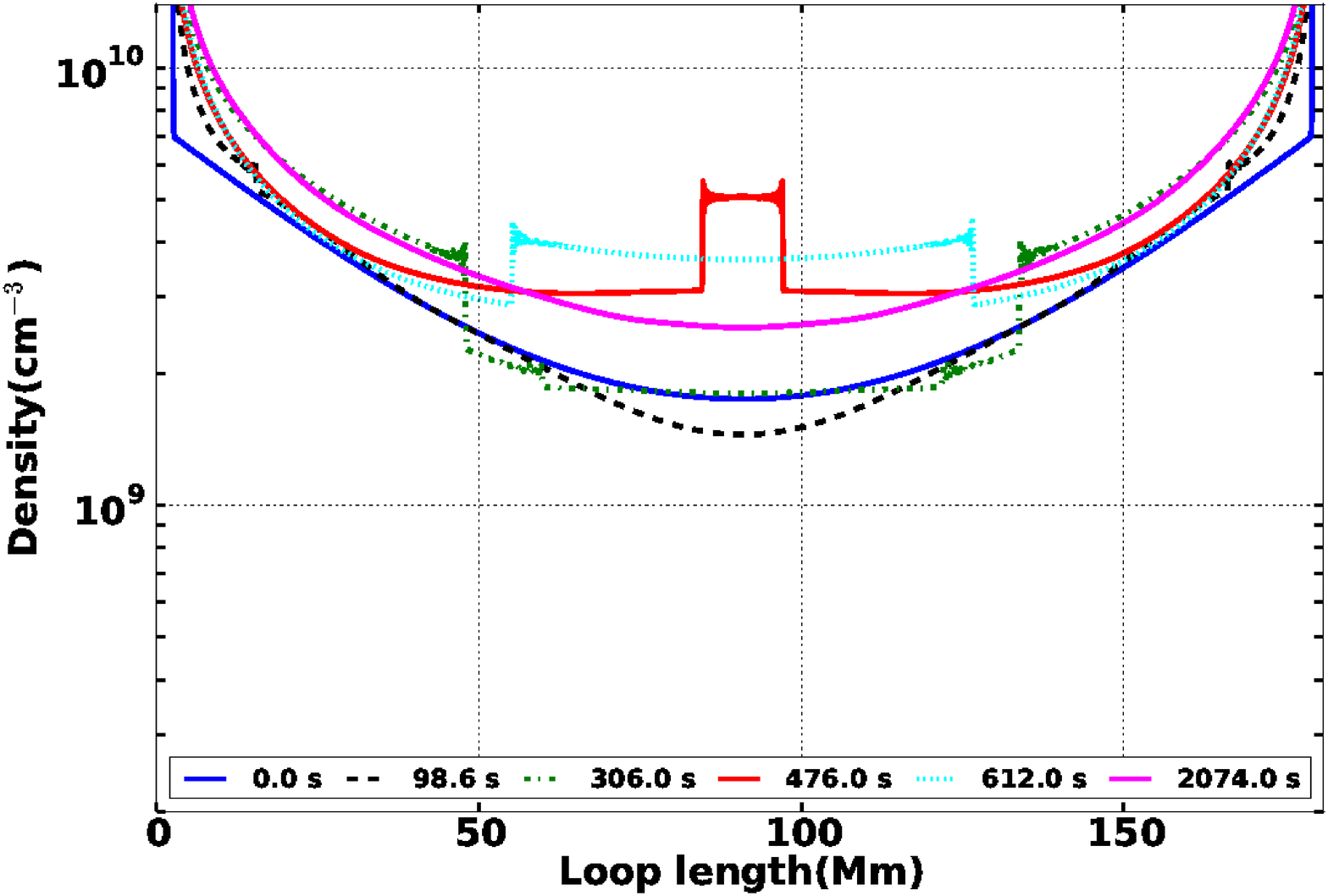}
  &
\includegraphics[scale=0.22]{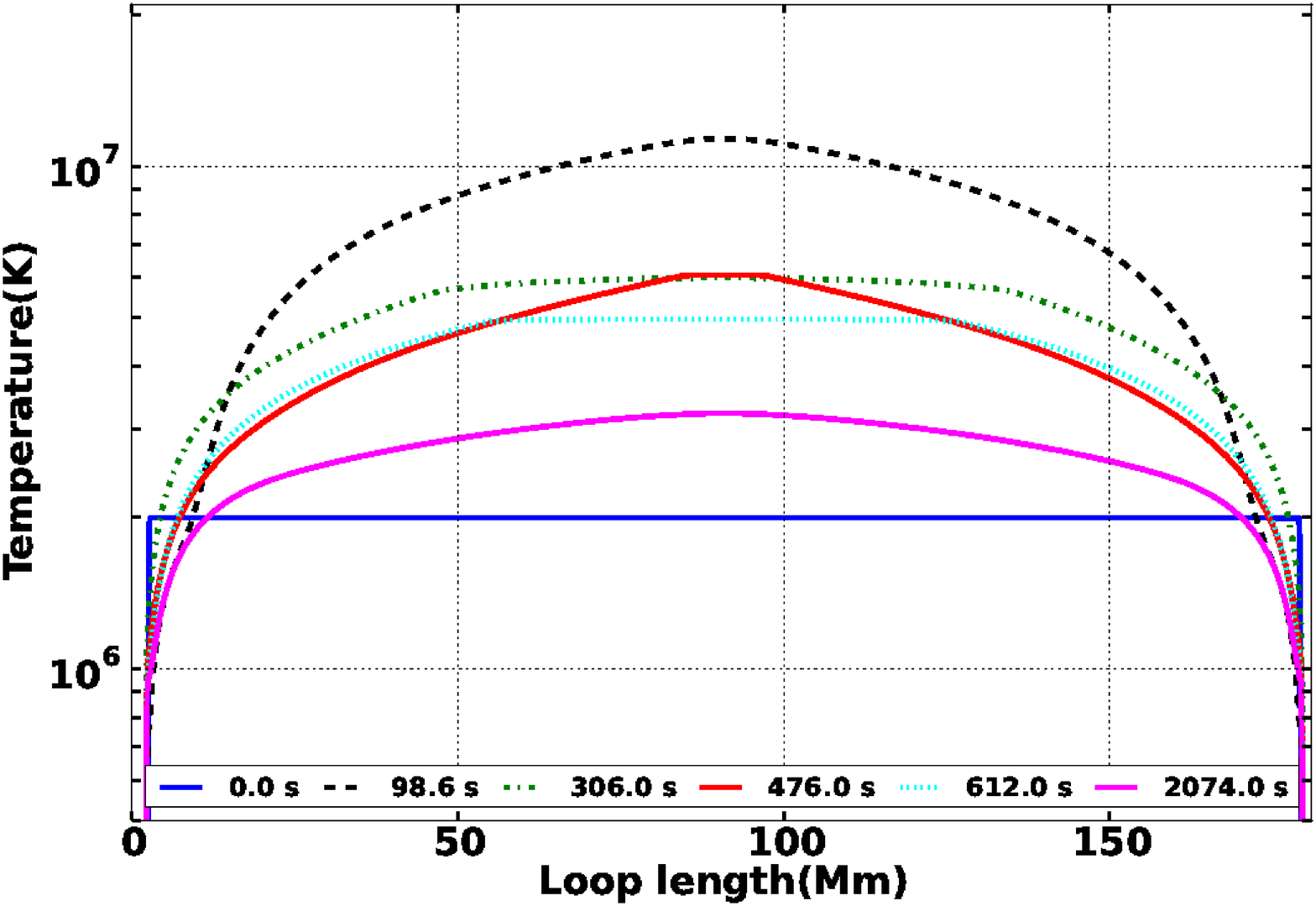}
 
 \\ (a) Density for 8 $\times$ 10$^{10}$ \ ergs/cm$^{2}$&(b)Temperature for 8 $\times$ 10$^{10}$ \ ergs/cm$^{2}$\\
\includegraphics[scale=0.22]{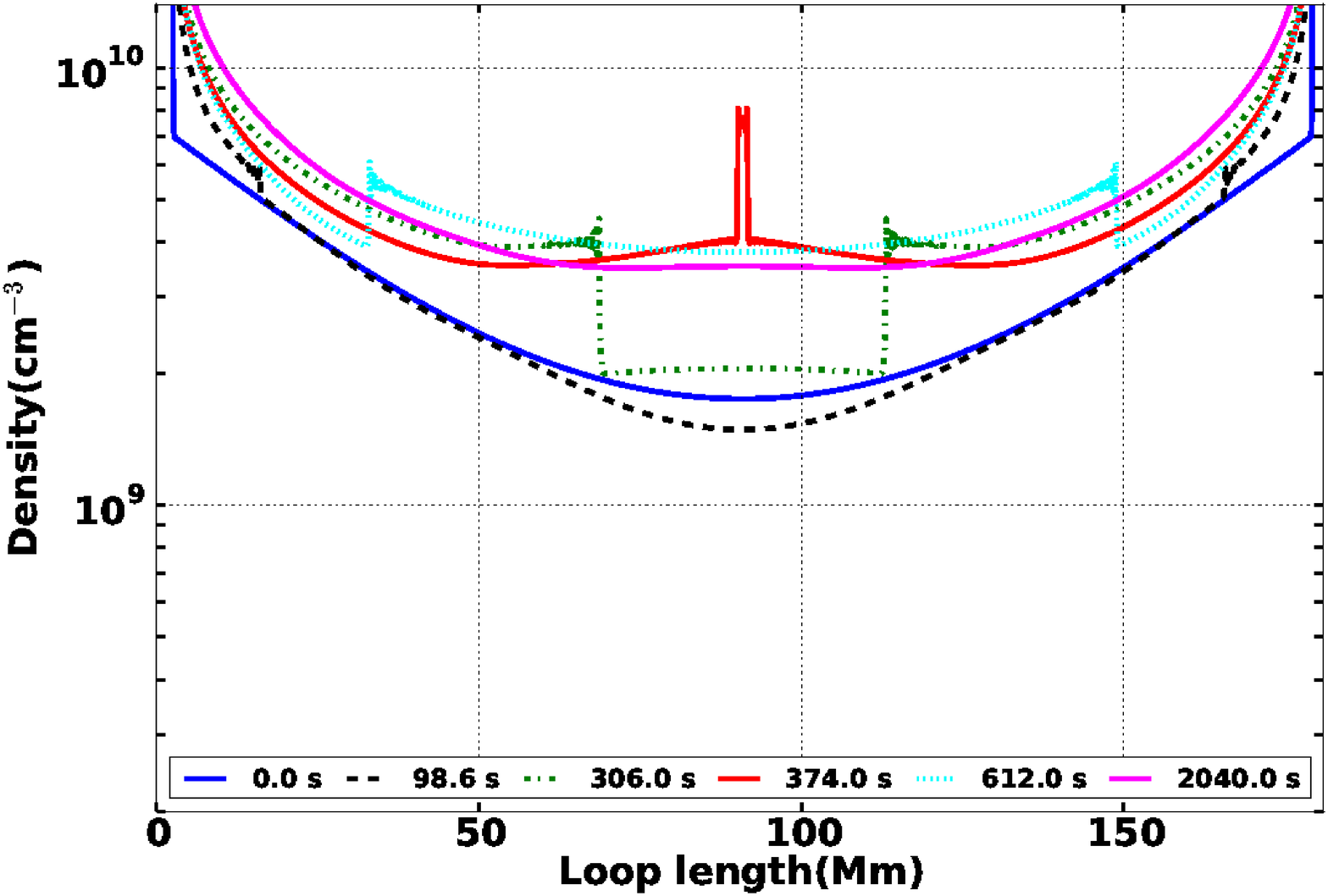}
 &
 \includegraphics[scale=0.22]{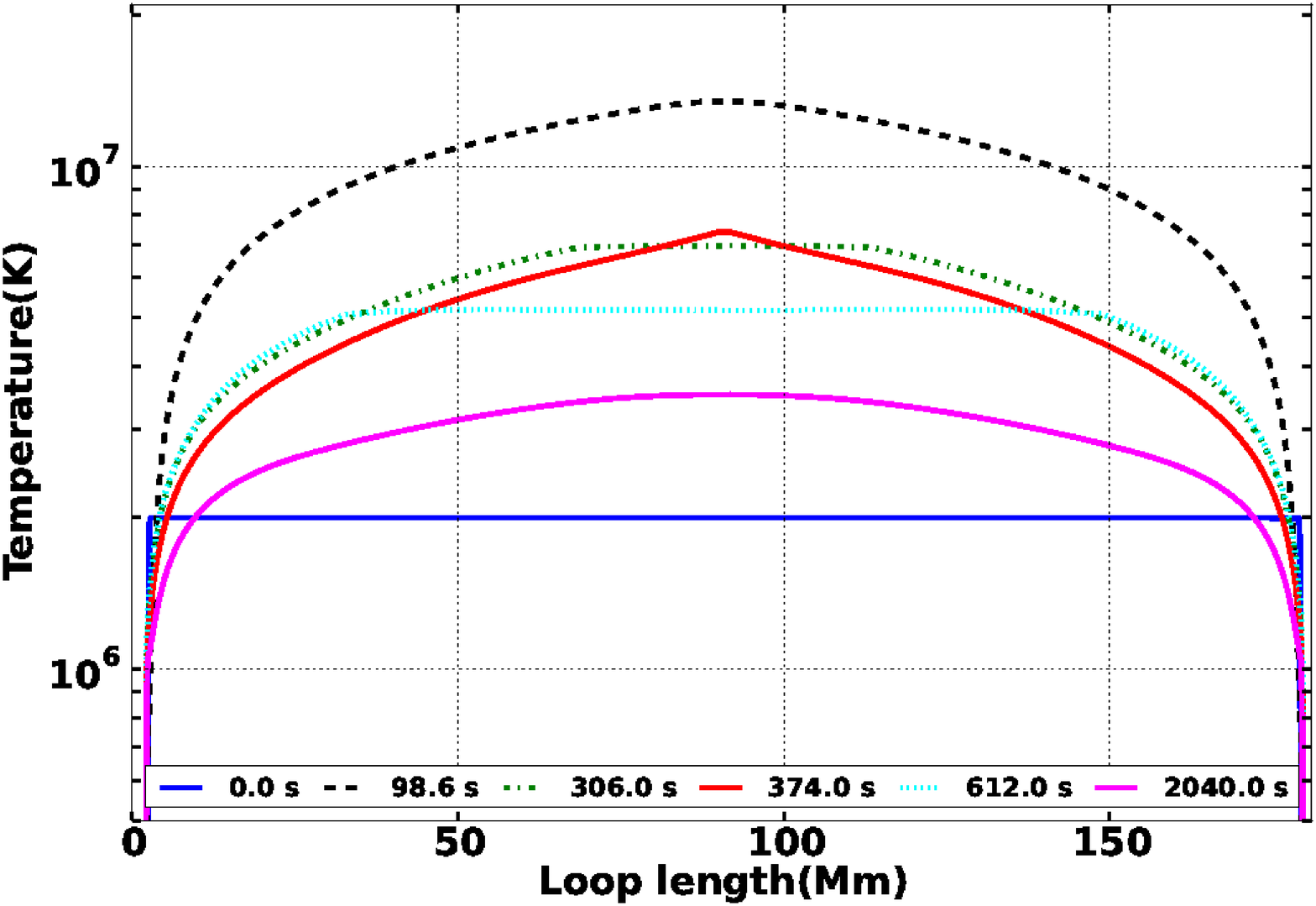}
 \\ (c)Density for 1.2 $\times$ 10$^{11}$ \ ergs/cm$^{2}$&(d)Temperature for 1.2 $\times$ 10$^{11}$ \ ergs/cm$^{2}$\\
 \includegraphics[scale=0.22]{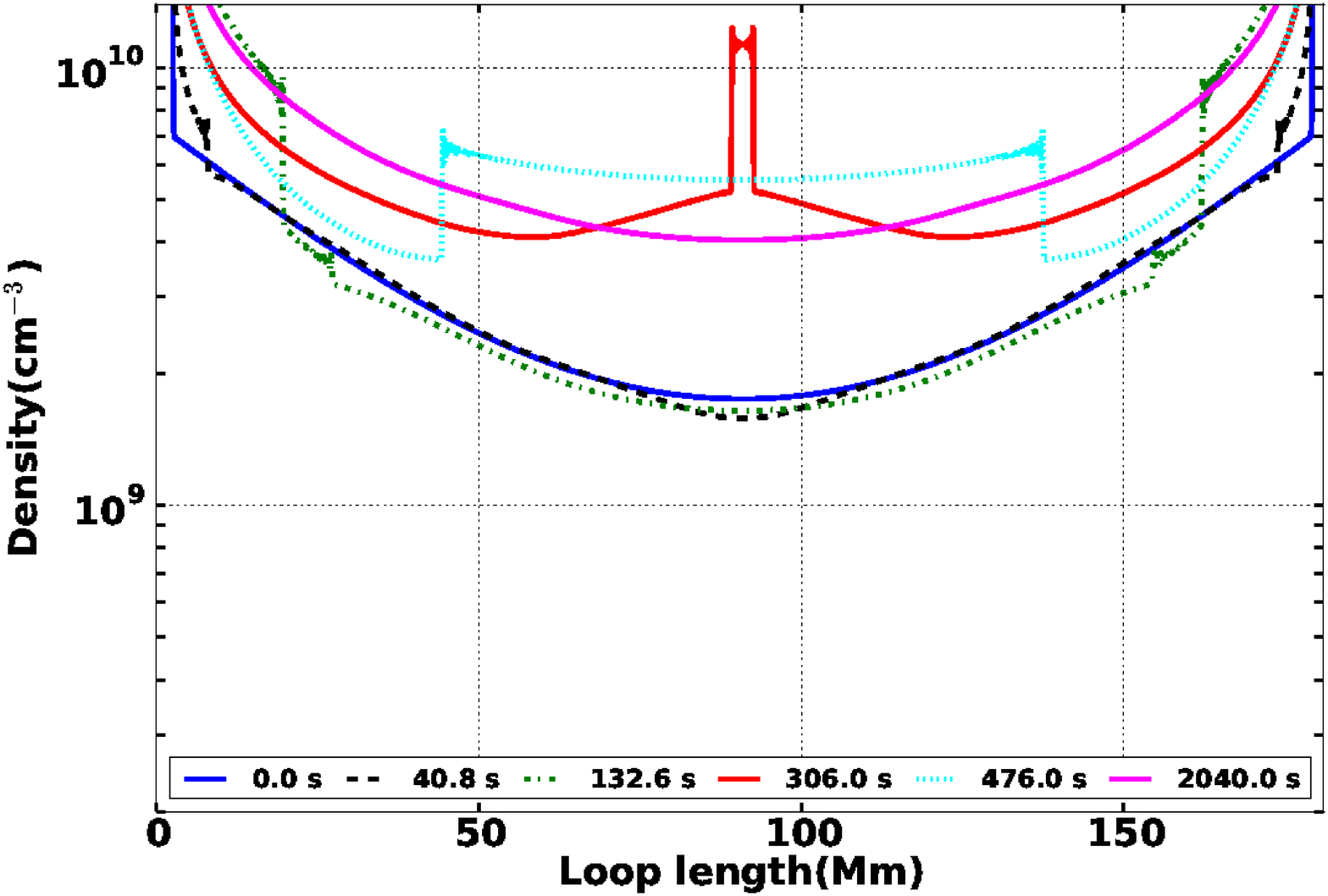}
 &
 \includegraphics[scale=0.22]{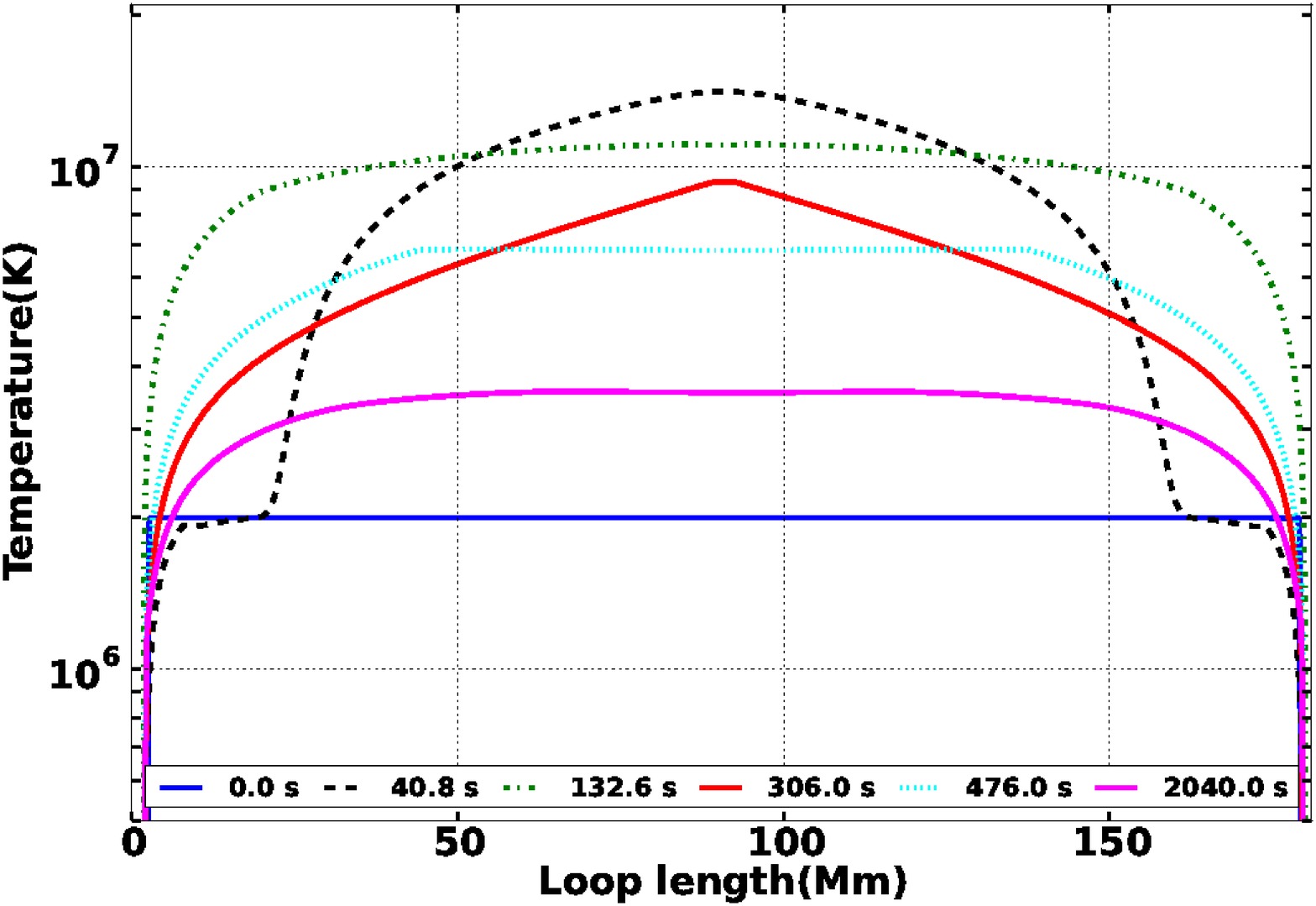}
 \\ (e)Density for 2 $\times$ 10$^{11}$ \ ergs/cm$^{2}$&(f)Temperature for 2 $\times$ 10$^{11}$ \ ergs/cm$^{2}$
\end{tabular}
\end{center}
\caption{Density and Temperature evolution for various heating strengths. \label{density_temp2}}
\end{figure*}

The time evolution of density and temperature along the loop for different heating strengths is shown in 
Figures~\ref{density_temp1} \& \ref{density_temp2}. The time steps shown are chosen so as to describe the collision of primary 
evaporation at the loop top. The strength of externally supplied heat is given in the figure. As can be clearly seen in 
the density plots of Fig.~\ref{density_temp1} \ref{density_temp2}, the primary flow collision creates an enhancement in density at the loop top. 
With the increasing strength of external heat, the enhancement in density increases and 
starts to appear at earlier times. The temperature profile of the corona increases uniformly for large heating inputs. To make a 
quantitative assessment in the changes in density and temperature with increasing heat input, we have defined a parameter "density 
increment factor" 
($\delta$) as the ratio of the loop top density after the collision to the ambient coronal density. 
The variation of $\delta$ and loop top temperature with increasing external heat input is given in Table~\ref{DandT}. The table 
clearly reveals that density increment factor $\delta$ as well as the loop top temperature increases with the increasing 
strength of external heat input. We also calculate the Mach number very close to the loop top just after the collision. For all heating strengths the primary collision remains subsonic.

%%----------------------------------------------------------------------
\begin{table}
\centering
\caption{Loop top parameters at the time of primary shock collision for various heating strengths. NOTE: The Mach number is calculated 
adjacent to the loop top. \label{DandT}}
\begin{tabular}{ccccc}
\tableline\tableline
\vbox{\hbox{\strut H$_{f}\times$ 10$^{9}$}\hbox{\strut(~ergs~cm$^{-2}$)}} & \vbox{\hbox{\strut Factor of }\hbox{Increment ($\delta$)}}  & \vbox{\hbox{\strut Coronal density}\hbox{($\times$ 10$^{8}$ cm$^{-3}$)}}& \vbox{\hbox{\strut \textbf{T$_{top}$}} \hbox{(MK)}} &\vbox{\hbox{\strut M$_{a}$}\hbox{}} \\
\tableline
$2.0 $ & 1.21& $1.52 $&1.97&0.24\\
$20.0 $ & 1.35 &$2.32 $&3.49&0.33\\
$40.0  $ & 1.36 &$2.48 $&4.73&0.39\\
$80.0  $ & 1.69 &$3.98 $&6.48 &0.60\\
$120.0 $ & 1.86 &$4.15$&7.61&0.63\\
$200.0 $ & 2.17 &$4.32 $&9.43&0.62\\
\tableline
\end{tabular}
\end{table}
%%----------------------------------------------------------------------
 
The increase in density and temperature at loop tops with increasing external heat input can be explained as following. The 
stronger heat input creates stronger conduction front from loop top towards the footpoint and dumps larger matter and energy in the chromosphere. This leads to a stronger evaporation flow. In all the case, the coronal temperature increases till the first 100~seconds corresponding to the heat input time and then drops as the loop cools. Later on, the coronal part of the loop near the loop 
top show an enhancement in the temperature which corresponds to the time of the primary collision of the evaporation flow. This temperature 
enhancement is strongly related to enhancement in the density (see Figures.~\ref{density_temp1} \& \ref{density_temp2}). After the 
collision of the primary chromospheric evaporation flow, the reverse flow is created that moves towards chromosphere and collides with 
secondary and tertiary flows coming from the chromospheric regions to coronal region. These flow collisions occur at various places in the 
loop depending on the flow speeds. However, these collisions are weak and the temperature  and density increment is less compared 
to the primary collision at the loop top. We expect subsequent reverse flow collisions to be strong for asymmetric heating cases. It may explain the observation of knots seen in EUV observations and modelled by \cite{patsourakos2004}.

%%----------------------------------------------------------------------
%%----------------------------------------------------------------------

%%----------------------------------------------------------------------
%%----------------------------------------------------------------------

%%----------------------------------------------------------------------
\section{Forward Modelling of Spectral lines Observed with EIS }\label{foward_modelling}
%%----------------------------------------------------------------------

The enhancement in the loop top density and temperature is expected to get reflected in various coronal emission lines. In this section, 
we forward model various spectral lines observed by EIS covering a range of peak formation temperature of 1.0 to 5.6~MK. The spectral lines which are 
forward modelled and their peak formation temperature is given in Table~\ref{lines}. The peak formation temperature and other atomic 
parameters are obtained from CHIANTI v.7.0 \citep{dere1997,landi2013}. 

%%----------------------------------------------------------------------
\begin{table}
\centering
\caption{The EIS spectral lines chosen for forward modelling \label{lines}}
\begin{tabular}{ccc}
\tableline\tableline
Ionised State  & Wavelength($\AA$) &  \vbox{\hbox{\strut Peak Formation }\hbox{Temperature (MK)}} \\
\tableline
Fe X & 184.54 &1.1\\
Fe XI & 188.23 &1.4\\
Fe XII  & 195.19&1.6\\
Fe XV & 284.16 &2.2\\
Ca XVII & 192.8 &5.6\\
\tableline
\end{tabular}
\end{table}
%%----------------------------------------------------------------------

Intensity of an optical thin emission line can be written as,

%%----------------------------------------------------------------------
\begin{equation}
I_{i,j}(\nu)= \int G_{ji}(T_{e},n_{e}) n_{e}^{2} dl
\end{equation}
%%----------------------------------------------------------------------

\noindent where i,j are upper and lower levels respectively. G($T_{e}$,n$_{e}$) is the contribution function of the spectral line that accounts for the ionisation fraction, 
coronal abundances (A$_{X}$). It has strong depends on electron temperature. The expression of G(T$_{e}$,n$_{e}$) is given by,

%%----------------------------------------------------------------------
\begin{equation}
G_{ji}(T,n_{e})=\frac{hc}{4 \pi \lambda_{ji}} \frac{A_{ji}}{n_{e}} \frac{N_{j}(X^{+k})}{N(X^{+k})}\frac{N(X^{+k})}{N(X)}A_{X} 
\end{equation}
%%----------------------------------------------------------------------

\noindent where A$_{ij}$ is the spontaneous transition probability for j to i transition. $\frac{N_{j}(X^{+k})}{N(X^{+k})}$ is the population of level j relative to the total
N(X$^{+k}$) number density of ion X$^{+k}$. $N(X^{+k})$ and $N(X)$ are the element species in k$^{th}$ ionised state and neutral state respectively. 

\begin{figure}[htbp]
\centering
\includegraphics[width=0.4\textwidth]{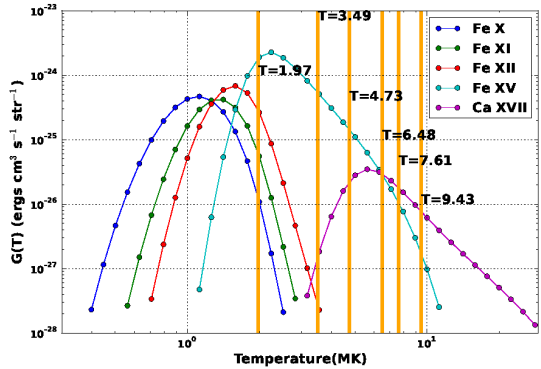}
\caption{Contribution function of the spectral lines listed in Table~\ref{lines} computed using a density of 10$^{9}$ cm$^{-3}$
and coronal abundances of Feldman et al. 1992 and CHIANTI. The orange lines corresponds to the loop top temperature (in MK) at the time 
of collision of primary flows corresponding to the six simulations with varying heart strength. The temperature increased from 
$T~=~1.97$~MK to $T~=~9.43$~MK corresponding to increasing heating strength. \label{contri}}
\end{figure}

We have computed the contribution function for each spectral lines using CHIANTI for a given electron density and coronal abundances 
of \cite{feldman1992}. Note that, in this study we are only forward modelling the spectral lines (given in Table~\ref{lines}) which are 
not density sensitive. The contribution functions for all the lines listed in Table~\ref{lines} are plotted in Figure~\ref{contri}.

We have forward modelled the loop intensities in all the spectral lines listed in Table~\ref{lines} using the density and temperatures
obtained at each point along the loop. We note that the forward modelling was only performed to the coronal part of the loop. This is 
essentially due to the fact that the assumption of optically thin atmosphere will breakdown near the footpoint of the loop in the chromosphere.

%%----------------------------------------------------------------------
%%----------------------------------------------------------------------

\begin{figure}
\centering
\includegraphics[width=0.4\textwidth]{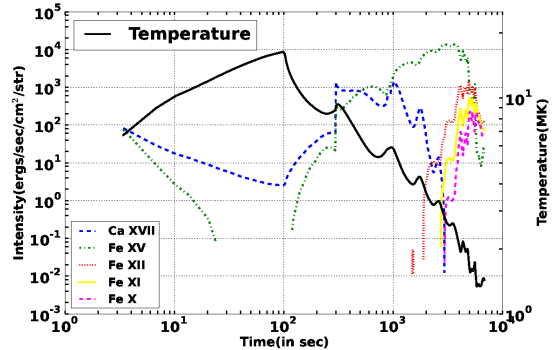}
\caption{The figure plots spectral line intensities and temperature of the loop top for external heating input of 2$\times$10$^{11}$~ergs~cm$^{-2}$. The bold solid line correspond to the temperature, dashed-dotted line is intensity of \ion{Fe}{15} lines, the dashed line is the intensity of \ion{Ca}{17} line, dotted line is the intensity of \ion{Fe}{12}, thin solid line is the intensity of \ion{Fe}{11} and thick dashed line is the intensity of \ion{Fe}{10}.\label{specific_intensity_loop_top}}
\end{figure}

Figure~\ref{specific_intensity_loop_top} displays the evolution of the intensities in these spectral lines at the loop top for the highest 
heating scenario. In the figure, the bold solid black line corresponds to the temperature at the top, dashed-dotted green line is intensity of \ion{Fe}{15} 
line, the dashed blue line is the intensity of \ion{Ca}{17} line, dotted red line is the intensity of \ion{Fe}{12}, thin solid yellow line is the intensity of 
\ion{Fe}{11} and thick dashed magenta line is the intensity of \ion{Fe}{10} line. The temperature at the loop top show increase till 100 sec, reflecting 
the time till which the external heating was switched-on and decreases afterwards. 

At initial times (t $\sim$ 3.4 sec), where the temperature is  
T $\sim$ 6.48~MK, the intensity of \ion{Fe}{15} and \ion{Ca}{17} are almost the same. As we see from Figure~\ref{contri}, coincidently, the contribution
functions of these two lines cross each other exactly at the temperature T =~6.48 MK that can explain the equal intensity. This temperature
nearly corresponds to the peak of the contribution function of \ion{Ca}{17} and lies in the far right side of that of \ion{Fe}{15}.
With the increase of temperature, the intensities in both the spectral lines fall. The emission in \ion{Fe}{15} completely disappears when the 
temperature reaches to roughly 12~MK, as this line has contribution function, albeit very minimal, reaching to a maximum temperature of 10~MK. 
The intensity in \ion{Ca}{17} also decreases but never disappears as the contribution function of this line goes as large as $\sim$~12~MK. 

After 100~seconds, i.e., when the external heating is switched-off, the loop top starts to cool and the intensity in \ion{Ca}{17} starts to increase, as 
we move towards the peak of it's contribution function. The emission from \ion{Fe}{15} also reappears, albeit later (at t= 101~sec). At 
t=300~sec, the temperature show a slight enhancement due to the primary collision of the evaporation flow (as explained in \ref{dynamics} and shown 
Figure~\ref{density_temp2}). Correspondingly, the intensities in \ion{Fe}{15} and \ion{Ca}{17} show a slight dip followed by a very 
sharp increase, which is about two orders of magnitude higher. This sharp enhancement in the intensities coincide with the time of the collision of 
the primary evaporation flow collision as well as the enhancement in the density. Later on, the temperature falls and shows many bumps before it
dies out. The intensity in \ion{Ca}{17} remains similar with some fluctuations till t=1000~s and falls sharply afterwards. Whereas, the intensity of 
\ion{Fe}{15} keeps increasing with oscillations all the way till t=3000~s that corresponds to a temperature of 3~MK. The oscillations in the intensities 
of \ion{Ca}{17} as well as in \ion{Fe}{15} corresponds well to the times of bumps see the temperature profile. These oscillations are essentially due 
to the collisions of the subsequent evaporation flows. The fall of intensities of these two lines corresponds to the cool-ward fall of their contribution 
function. The spectral lines \ion{Fe}{12}, \ion{Fe}{11} and \ion{Fe}{10} appears at times t=1040, 1150, 1180~sec respectively and they do not fade away 
till the end of simulation. As the loop cools, the other spectral lines gain in the contribution function and results in the enhancement in their 
intensities. Also note that the appearance of \ion{Fe}{12} preceded that of \ion{Fe}{10} appears, which can be attributed to their corresponding
contribution function.

%%----------------------------------------------------------------------

%%----------------------------------------------------------------------

After the primary collision (t=304~sec), the dips in the temperature correlates with the peaks in the intensities of \ion{Fe}{15}, \ion{Fe}{12}, 
\ion{Fe}{11} and \ion{Fe}{10}. Whereas, reverse is true for \ion{Ca}{17} as temperature dips correlates with the \ion{Ca}{17} intensity dips. This is essentially due to the fact that the temperature of the loop 
top is less than 5.6~MK (peak formation temperature of \ion{Ca}{17}) and it falls on the positive slope of the contribution function of \ion{Ca}{17} and
negative slope of that of \ion{Fe}{15}, \ion{Fe}{12}, \ion{Fe}{11} and \ion{Fe}{10} lines (see Figure~\ref{contri}). If temperature lies in negative slope in G(T) curve, any increase in the temperature should show decreases in the intensities of these lines and vice versa. However, we can see in Figures \ref{density_temp1} and \ref{density_temp2} , the collision of evaporation flows also enhances the 
densities that could in turn enhance the emission measure thereby showing the increase in the intensities. The factor of increment of densities $\delta$ for this largest heating case is 2.17 seen in Table \ref{DandT}.   

\begin{figure*}
\begin{center}
\begin{tabular}{cc}

\includegraphics[scale=0.20]{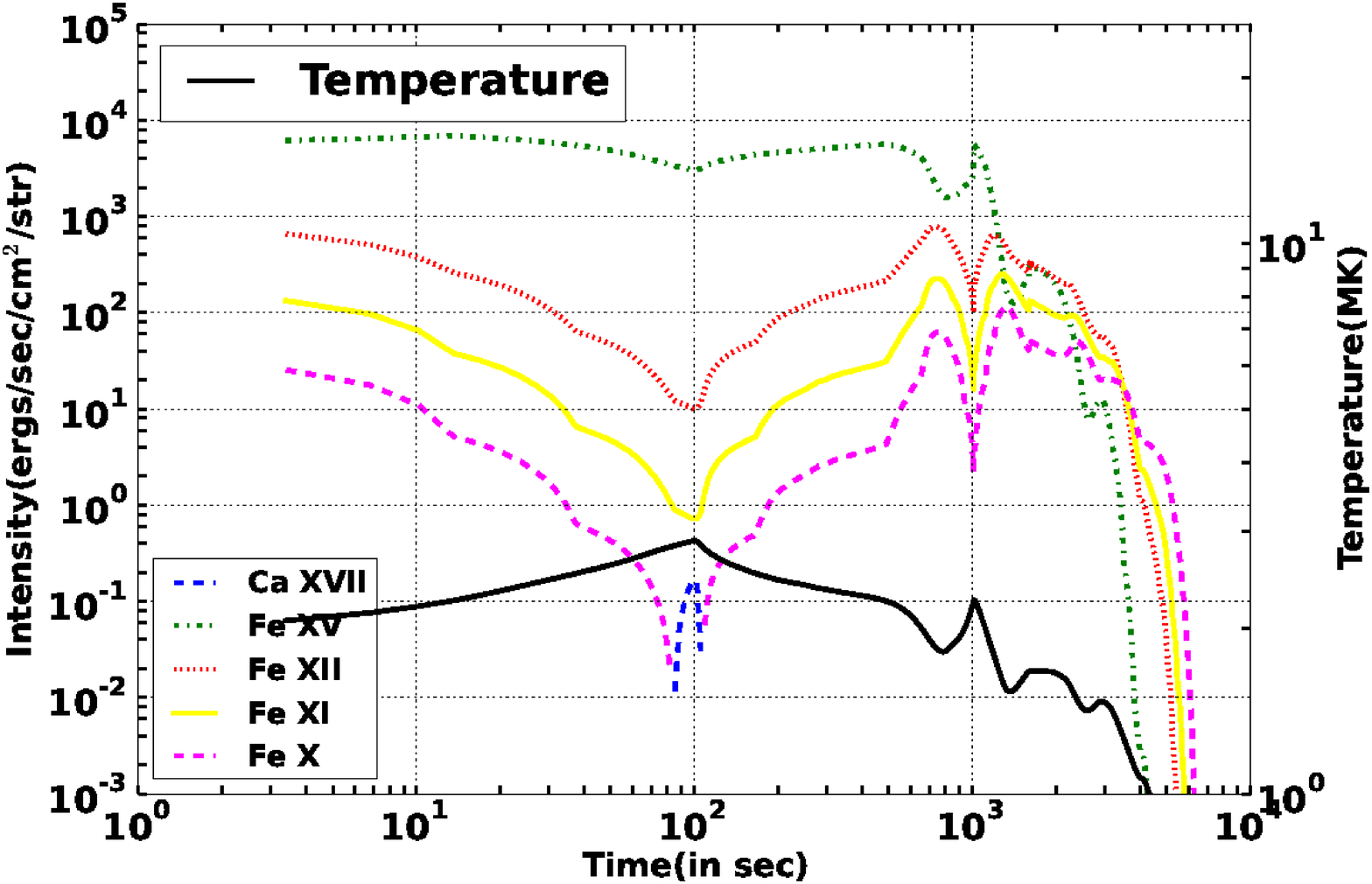}
  &
\includegraphics[scale=0.20]{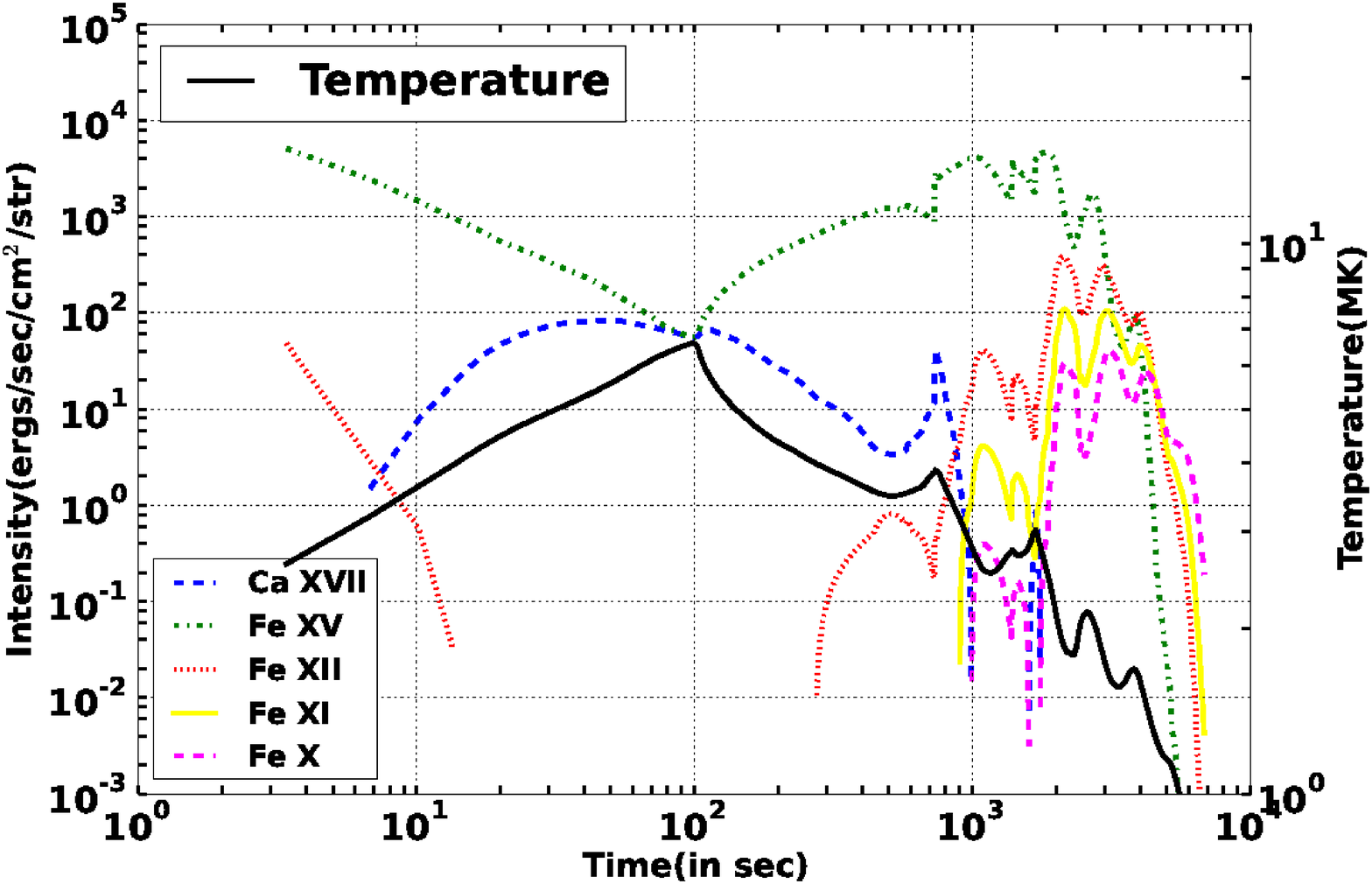}
 \\
 (a) 2 $\times$ 10$^{9}$ \ ergs/cm$^{2}$&(b) 2 $\times$ 10$^{10}$ \ ergs/cm$^{2}$ \\
\includegraphics[scale=0.20]{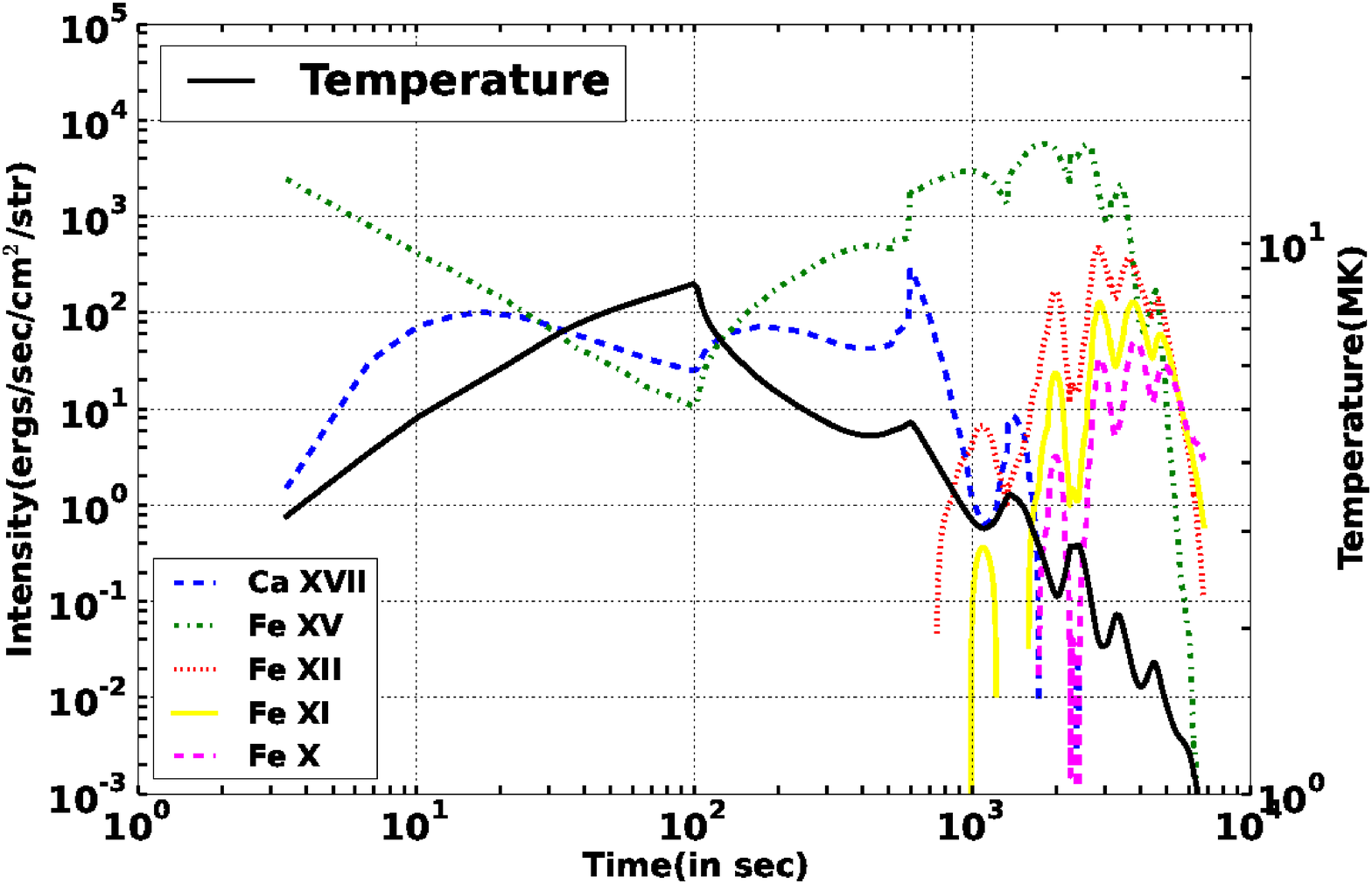}
 &
\includegraphics[scale=0.20]{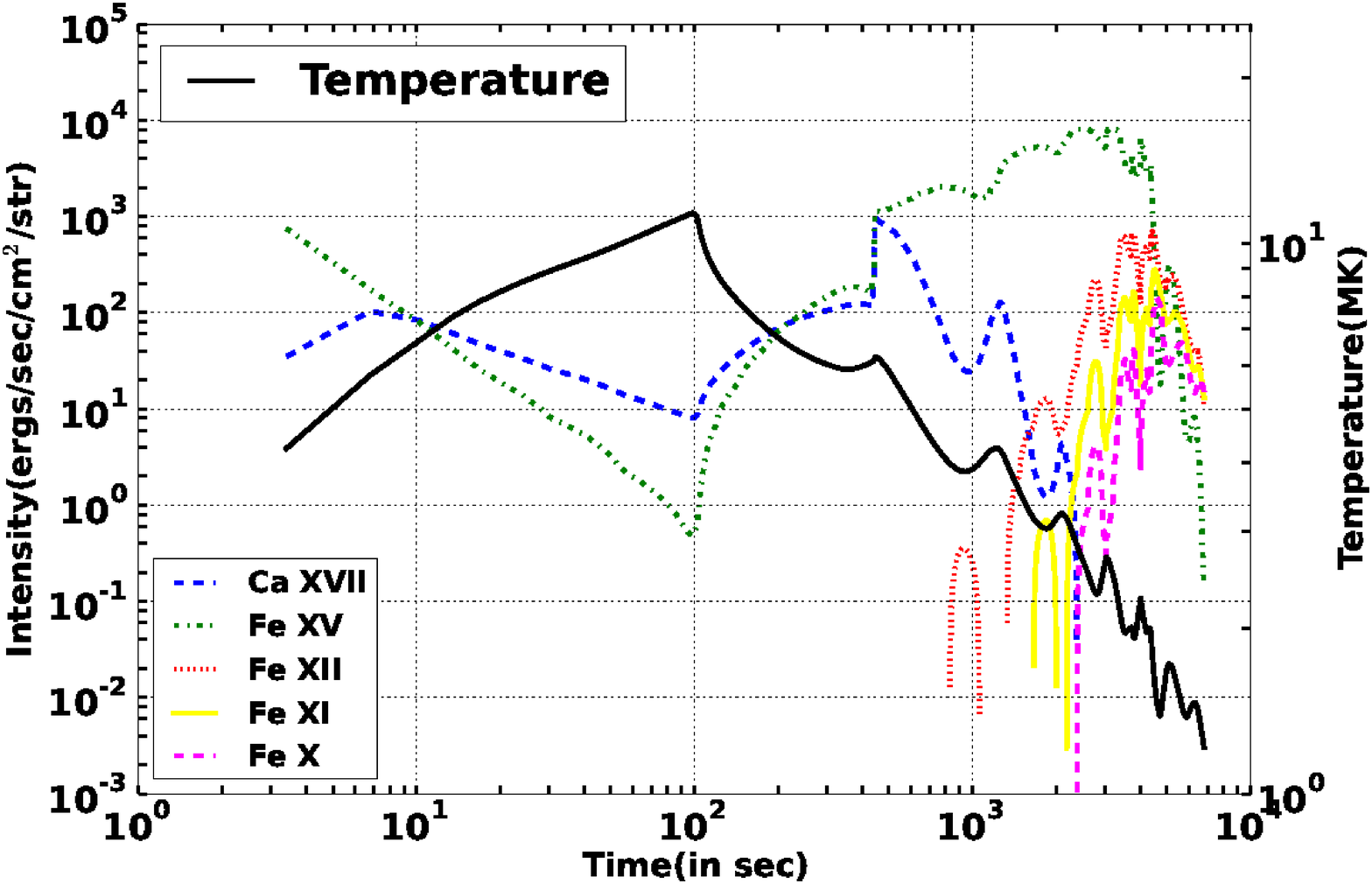}
\\
(c) 4 $\times$ 10$^{10}$ \ ergs/cm$^{2}$ & (d) 8 $\times$ 10$^{10}$ \ ergs/cm$^{2}$
\\
\includegraphics[scale=0.20]{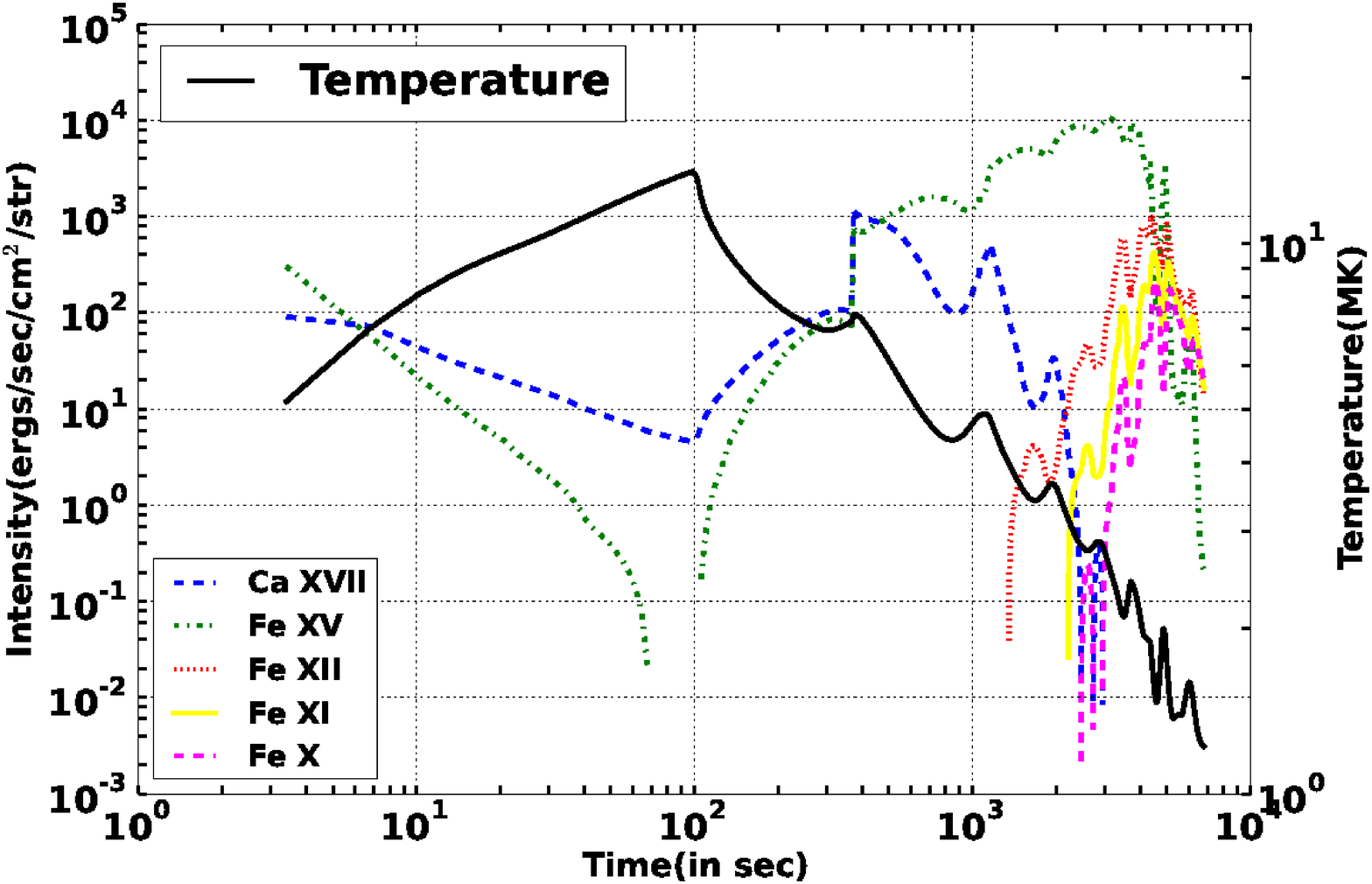}
 &
\includegraphics[scale=0.20]{fig9.eps}
  \\
  (e) 1.2 $\times$ 10$^{11}$ \ ergs/cm$^{2}$&(f) 2 $\times$ 10$^{11}$ \ ergs/cm$^{2}$

\end{tabular}
\end{center}
\caption{The panel shows the plots of the intensity variation of the loop top with time for different spectral lines along with the temperature of the loop top.}
\label{loop_top}
\end{figure*}
%%----------------------------------------------------------------------

Figure~\ref{loop_top} displays the evolution of intensities at the loop top for different heating strengths. In the figure, the bold black solid line 
correspond to the temperature of the top, dashed-dotted green line is intensity of \ion{Fe}{15} lines, the dashed blue line is the intensity of \ion{Ca}{17} 
line, dotted red line is the intensity of \ion{Fe}{12}, thin solid yellow line is the intensity of \ion{Fe}{11} and thick dashed magenta line is the intensity of 
\ion{Fe}{10}. While explaining the evolution of loop top emission in various lines, similar argument will follow as is given earlier for the highest 
heating case. As we can see, the appearance and disappearance of the emission in various spectral lines corresponds well with their respective 
contribution functions. For example, for the lowest heating case, the emission in \ion{Ca}{17} appears for a very short duration, only at the time
of peak temperature achieved during heating. Afterwards, the emission on \ion{Ca}{17} dies down. Similarly, the \ion{Fe}{10} line appears strongly for 
the lower heating case much earlier than it appears in the higher heating case (see Figure \ref{loop_top}).

We have provided the density increment factor ($\delta$) due to primary collision of the evaporation flow in Table~\ref{DandT}. Corresponding increment in the forward modelled intensities ($\beta_{sim}$) in various spectral lines are given in Table~\ref{table_int}. Similar to the definition of density increment factor, the intensity increment factor ($\beta_{sim}$) is also defined as the ratio of intensities at the loop top followed by the primary collision and to the ambient coronal intensity. 

The intensity increment factor depends on the temperature of the loop top and densities. As density increment factor ($\delta$) increases with input heat(Table \ref{DandT}), we obtain larger $\beta_{sim}$ for all spectral lines (Table \ref{table_int}). So in all times, the density become most important in increasing loop top intensity at the time of primary collision. The temperature of the loop top also decides the intensity increment or decrement as seen in Fig \ref{loop_top} depending on the position on the contribution function (Fig \ref{contri}). 

As can be seen from the table, the loop top temperature can reach up to $\sim$~10~MK for \ion{Ca}{17} for highest heating. It created the maximum intensity increment of about a factor of 4.4 for \ion{Ca}{17}. The typical $\beta_{sim}$ is in the range $\sim~1.5-3.5$, considering all heating strengths and are close to brightness factor in observation $\beta_{obs}$ (Table \ref{table1}). But among all heating inputs, the 8.0 $\times$ 10$^{10}$ ergs cm$^{-2}$ input heat case's loop top increment $\beta_{sim}$ shows the closest resemblance with the  $\beta_{obs}$ for all spectral lines, except \ion{Fe}{11}. The case of 1.2 $\times$ 10$^{11}$ ergs cm$^{-2}$ is closest to \ion{Fe}{11}. Hotter lines produces larger increments $\beta_{sim}$ than cooler lines and it increases as input heat increases. In Table \ref{table1}, this trend does show up from \ion{Fe}{13} to \ion{Ca}{17}. In simulations the \ion{Fe}{15} line (Fig. \ref{loop_top}, dashed green curve) is dominant for all heating strengths in post-flare times, which is in agreement with the loop top intensity of the observation (Table \ref{table1}). We note here that there may not be any one-to-one correlation between the observations and forward modelled results. This could be due to the limitations of the simulations and forward modelling. In the simulations, the dynamics corresponding to one loop strand is being studied, whereas in simulations there will be a collection of strands going through heating and cooling phases simultaneously.

%%----------------------------------------------------------------------
\begin{table}
\centering
\caption{The factor of increment ($\beta_{sim}$) in the modelled EIS intensities for various spectral lines for different heating strengths along with the temperature of the loop top.
\label{table_int}}
\begin{tabular}{|c|c|c|c|c|c|c|}
\hline
\vbox{\hbox{H$_{f}\times 10^{9}$}\hbox{(ergs~cm$^{-2}$)}}&Fe X&Fe XI&Fe XII&Fe XV&Ca XVII&\vbox{\hbox{T$_{top}$}\hbox{(MK)}}\\

\tableline
$2.0 $ &1.15&1.22&1.30&1.64&2.59&1.97\\
$20.0$ &1.52&1.55&1.60&1.76&2.12&3.49\\
$40.0$ &1.67&1.55&1.61&1.76&1.94&4.73\\
$80.0$ &3.20&2.81&2.77&2.81&2.84&6.48 \\
$120.0$ &3.24&1.64&3.31&3.38&3.42&7.61\\
$200.0$ &5.06&6.45&2.79&3.83&4.42&9.43\\
\tableline

\end{tabular}
\end{table}
%%----------------------------------------------------------------------
%%----------------------------------------------------------------------
\section{Summary and Discussion }\label{summary}
%%----------------------------------------------------------------------

In this paper we have addressed the long standing problem of localised loop top brightening in post flare loops observed in soft X-ray and EUV. We performed a 1D hydrodynamic simulation and forward modelled the intensities in a few spectral lines (see Table~\ref{lines}) observed by EIS. Our results show that the external heating applied at the loop top produces evaporations flows form chromosphere (Fig. \ref{density_temp1} \& \ref{density_temp2}). These flows collide at the loop top and enhance the density at the top. This can be clearly seen as increase in the intensities in various spectral lines (Fig. \ref{loop_top}). This idea of density enhancement in coronal loops due to chromospheric evaporation is also supported in recent studies such as \cite{takasao2015} and reference therein. Moreover, once the heating is switched-off, the temperature of the loop top drops sharply due to enhanced radiative cooling. The decline in temperature is larger for stronger heating cases. This is essentially due to the fact that evaporation flows are stronger bringing more mass in the loops thereby increasing the density. We find that temperature of the loop top is important parameter which determines loop top intensity in post flare times. However, the density increments $\delta$ (see Table \ref{DandT}) are also proportional to the strength of loop top intensities (Table \ref{table_int}). The intensity increments ($\beta_{sim}$) at the time of primary collision are close to observations $\beta_{obs}$ (Table \ref{table1}). Our forward modelling shows that  \ion{Fe}{15} to the strongest lines for all the heating strengths in the post flare loops. This is consistent with the observations provided by \cite{Hara2006} as shown in Table \ref{table1}. In post flare times, the cooler spectral lines \ion{Fe}{12},\ion{Fe}{11} and \ion{Fe}{10} appears in the decreasing order of their formation temperatures. 

 Finally, we conclude that larger external heating produces stronger the evaporation and therefore, creates stronger the loop top brightening in hotter lines. Therefore, the post flare loops formed in higher energetic flare cases will appear brighter. We know that most of the strong flares are observed from complex magnetic regions, for example active regions. However, the low energy flares are often observed from quiescent part of sun's disk. In these cases, a post flare loop formed due to flare eruption of a subsequent CMEs may not be bright enough to be observed. Further investigations regarding effects of various parameters such as heating locations, geometry of the loops as well the length and cross sections are required and is under progress. In addition, more observational study are required to provide quantitative constrains on the relative brightening in different spectral lines.

\acknowledgements{We thank the referee for careful reading and providing the constructive comments. RS acknowledges the support of Dr Divya Oberoi, NCRA-TIFR, Pune. DT and AG acknowledges the Max-Planck Partner Group of MPS at IUCAA. Hinode is a Japanese mission developed and launched by ISAS/ JAXA, collaborating with NAOJ as a domestic partner, NASA and STFC (UK) as international partners. Scientific operation of the Hinode mission is conducted by the Hinode science team organized at ISAS/JAXA. This team mainly consists of scientists from institutes in the partner countries. Support for the post-launch operation is provided by JAXA and NAOJ (Japan), STFC (UK), NASA, ESA, and NSC (Norway). CHIANTI is a collaborative project involving George Mason University, the University of Michigan (USA) and the University of Cambridge (UK). The authors acknowledge the CANS (Coordinated Astronomical Numerical Software)}.

\bibliography{manuscript}
\end{document}